\documentclass[12pt]{article}
\usepackage[flushleft]{threeparttable}
\usepackage{booktabs}
\usepackage{float}
\usepackage{natbib}
\usepackage{amsmath,amssymb,amsfonts}
\usepackage{graphicx,color,graphics,caption,epsf,dcolumn}
\usepackage{epstopdf}
\usepackage{amssymb,cases}

\begin{document}
\newcommand{\beq}{\begin{equation}}
\newcommand{\eeq}{\end{equation}}

\title{\Large \bf  The water vapor budget of a hurricane as dependent on its movement}

\author{Anastassia M. Makarieva$^{1,2}$\thanks{\textit{Corresponding author.} {E-mail: ammakarieva@gmail.com}},
Victor G. Gorshkov$^{1,2}$,\\ Andrei V. Nefiodov$^{1}$,
 Alexander V. Chikunov$^{3,4}$, Douglas Sheil$^{5}$,\\ Antonio Donato Nobre$^{6}$, and Bai-Lian Li$^{2}$}

\date{\vspace{-5ex}}

\maketitle

$^{1}$Theoretical Physics Division, Petersburg Nuclear Physics Institute, 188300 Gatchina, St.~Petersburg, Russia.
$^{2}$USDA-China MOST Joint Research Center for AgroEcology and Sustainability, University of California, Riverside 92521-0124, USA.
$^{3}$Institute of World Ideas, Udaltsova street 1A, 119415 Moscow, Russia.
$^{4}$Princeton Institute of Life Sciences, Princeton, USA.
$^{5}$Norwegian University of Life Sciences, \AA s, Norway.
$^{6}$Centro de Ci\^{e}ncia do Sistema Terrestre INPE, S\~{a}o Jos\'{e} dos Campos SP 12227-010, Brazil.

\begin{abstract}
Despite the dangers associated with tropical cyclones and their rainfall, the origins of storm moisture remains unclear.
Existing studies have focused on the region $40$-$400$ km from the cyclone center. It is known that the rainfall within this area
cannot be explained by local processes alone but requires imported moisture.  Nonetheless, the dynamics of this imported moisture
appears unknown. Here, considering a region up to three thousand kilometers from storm center, we analyze precipitation,
atmospheric moisture and movement velocities for North Atlantic hurricanes. Our findings indicate
that even over such large areas a hurricane's rainfall cannot be accounted for by concurrent evaporation.
We propose instead that a hurricane consumes pre-existing atmospheric water vapor as it moves. The propagation velocity of the cyclone, i.e.
the difference between its movement velocity and the mean velocity of the surrounding air (steering flow), determines the water vapor budget.
Water vapor available to the hurricane through its movement makes the hurricane self-sufficient at about $700$~km from the hurricane center
obviating the need to concentrate moisture from greater distances. Such hurricanes leave a dry wake, whereby rainfall is suppressed
by up to 40\% compared to its long-term mean. The inner radius of this dry footprint approximately coincides with the radius of hurricane
self-sufficiency with respect to water vapor. We discuss how Carnot efficiency considerations do not constrain the power
of such open systems that deplete the pre-existing moisture. Our findings emphasize the incompletely understood
role and importance of atmospheric moisture supplies, condensation and precipitation in hurricane dynamics.
\end{abstract}

\section{Introduction}

Intense tropical cyclones,  also called hurricanes and typhoons depending on their locations,
are associated with intense rainfall \citep{rodgers94,lonfat04}.
For example, the mean rainfall within an Atlantic hurricane
of 400~km radius is about 2 mm~h$^{-1}$. Such rainfall is ten times the long-term mean of tropical oceanic evaporation
and could deplete local atmospheric moisture within a day. However, hurricanes can sustain this rainfall for several days, during
which their atmospheric moisture content remains constant or even grows.
How do hurricanes achieve that?

Most empirical and modelling studies investigating the water budget of tropical cyclones have
focused on the inner area within $400$~km from the cyclone center. For Atlantic hurricanes
$r_o = 400$~km corresponds to the mean radius of the outermost closed isobar.
Within this area the local source of moisture (evaporation from the ocean) was estimated to be about
$10$-$20$\% of total rainfall (Table~1). This means that most moisture is imported from the outer region (Fig.~\ref{hur}).

Using the assumption that the cyclone is sustained by concurrent evaporation, many theoretical studies have viewed tropical
cyclones as steady-state thermodynamic systems which persist as long as they can consume moisture from the ocean
\citep{emanuel86,pauluis11,sabuwala15,ozawa15,kieu15}. But,
while oceanic evaporation within the cyclone area $r \leqslant r_o$ increases with wind speed and thus
exceeds its long-term mean value, the available estimates show that this is insufficient to sustain rainfall (Table~1).

On the other hand, long-term mean oceanic evaporation concurrent with the cyclone {\it could} provide sufficient water vapor to
sustain a high rainfall hurricane if this vapor were imported from
a large area of $1500$-$2000$~km around the hurricane center \citep{anthes74,trenberth03}.
This radius is $3$-$5$ times larger than that of the hurricane rainfall area. To concentrate
this moisture requires a radial pressure gradient to drive moist air towards the
storm's center. However, the radial pressure gradient in tropical cyclones is known to decline to near-zero values
at distances of a few hundred kilometers from the storm's center \citep[e.g.,][]{holland08}.
There are neither empirical nor theoretical evaluations of the ability of such pressure gradients to drive sufficient moisture
convergence across several thousand kilometers. Thus, despite the dangers associated with tropical cyclone rainfall
and the need for improved predictions \citep{rapp00}, there is no clarity as to how this rainfall is
generated and maintained.

Here we consider moisture dynamics in North Atlantic hurricanes up to
3000 km from the hurricane center. We first present theoretical considerations
which explain observed evaporation-to-rainfall ratios (Section~\ref{sec2}). These considerations demonstrate that observations do not support
the notion that hurricane-related rainfall is provided by evaporation concurrent with the hurricane, however large the spatial scale.
We then use data from the Tropical Rainfall Measuring Mission (TRMM) \citep{huffman07}
and Modern Era Retrospective Re-Analysis for Research and Applications (MERRA) \citep{rien11} to study rainfall and atmospheric moisture content
in the vicinity of selected hurricanes (Sections~\ref{met} and \ref{sig}).

Based on this analysis, we formulate a novel view of hurricane water budgets (Section~\ref{mot}). We show that the water vapor expenditures of
a hurricane can be explained by considering its large-scale motion: the hurricane depletes pre-existing atmospheric water vapor as it moves
through the atmosphere. In this manner the hurricane becomes self-sustaining for water vapor over a region of just about $700$~km from its center.
In the frame of reference co-moving with the storm there is a flow of moist ambient air towards the hurricane. This flow, which is
proportional to the velocity of hurricane's movement, delivers water vapor to the hurricane obviating the need to concentrate moisture
from far outside.

We discuss the implications of our findings for understanding hurricanes, their rainfall and their power. We emphasize that the
processes by which hurricanes and other cyclonic storms obtain their moisture, either by concurrent evaporation or by depletion of pre-existing
stores of water vapor, are key for finding constraints on
their intensity (Section~\ref{eff}). In the concluding section we discuss perspectives for further research.

\begin{table}[tbh!] \small
\begin{threeparttable}
\caption{\label{table1}Available estimates of mean precipitation $\overline{P}$, evaporation $\overline{E}$
and their ratio within an area of up to $r$ from the center of tropical cyclones (ordered by increasing $r$).
Asterisks denote North Atlantic hurricanes.}
\vspace{0.2 cm}
\begin{tabular}{l c c c c c c l c}\toprule
\multicolumn{1}{c}{Storm } & \multicolumn{1}{c}{$r$}    &  \multicolumn{1}{c}{$\overline{E}/\overline{P}$} &
\multicolumn{1}{c}{$\overline{P}\vphantom{\overline{P}^1}$} &  \multicolumn{1}{c}{$\overline{E}$} &  \multicolumn{1}{c}{Date} & Ref.\\
 &(km)&  \multicolumn{1}{c}{} &  \multicolumn{1}{c}{(mm~h$^{-1}$)} &  \multicolumn{1}{c}{ (mm~h$^{-1}$)} &      \\
\midrule
*Helen	&    \hphantom{0}37	&	0.14\hphantom{0}	&	18\hphantom{.0}\hphantom{0}	&	2.6\hphantom{0}	&	1958/09/26	&	{\it 1}\\
Nari	&    \hphantom{0}50	&	0.055	&	22.3\hphantom{0}	&	1.22	&	2001/09/19	&	{\it 2}\\
*Bonnie	&    \hphantom{0}70	&	0.05\hphantom{0}	&	17.5\hphantom{0}	&	0.88	&	1998/08/22	&	{\it 3}\\
*Helen	&    \hphantom{0}74	&	0.14\hphantom{0}	&	18.6\hphantom{0}	&	2.6\hphantom{0}	&	1958/09/26	&	{\it 1}\\
*Katrina&	100	&	0.11\hphantom{0}	&	19.1\hphantom{0}	&	2.1\hphantom{0}	&	2005/08/28	&	{\it 4}\\
*Ivan	&    100	&	0.078	&	13.4\hphantom{0}	&	1.04	&	2004/09/14	&	{\it 4}\\
Fay	&    100	&	0.059	&		&		&	1998/08/15	&	{\it 5}\\
Fay	&    100	&	0.33\hphantom{0}	&		&		&	1998/08/14	&	{\it 5}\\
*Helen	&    111	&	0.17\hphantom{0}	&	13.9\hphantom{0}	&	2.4\hphantom{0}	&	1958/09/26	&	{\it 1}\\
*Daisy	&    150	&	0.18\hphantom{0}	&	\hphantom{0}3.3\hphantom{0}	&	0.6\hphantom{0}	&	1958/08/27	&	{\it 6}\\
*Daisy	&    150	&	0.12\hphantom{0}	&	\hphantom{0}8.6\hphantom{0}	&	1.0\hphantom{0}	&	1958/08/25	&	{\it 6}\\
Nari	&    150	&	0.096	&	\hphantom{0}6.55	&	0.63	&	2001/09/19	&	{\it 2}\\
*Bonnie	&    200	&	0.08\hphantom{0}	&	10.1\hphantom{0}	&	0.81	&	1998/08/22	&	{\it 3}\\
Morakot	&    240	&	0.088	&	\hphantom{0}8.3\hphantom{0}	&	0.73	&	2009/08/07	&	{\it 7}\\
Fay	&    300	&	0.17\hphantom{0}	&	\hphantom{0}1.4\hphantom{0}	&	0.24	&	1998/08/15	&	{\it 5}\\
Fay	&    300	&	0.29\hphantom{0}	&	\hphantom{0}0.7\hphantom{0}	&	0.2\hphantom{0}	&	1998/08/14	&	{\it 5}\\
*Katrina	&  400	&	0.26\hphantom{0}	&	\hphantom{0}4.3\hphantom{0}	&	1.1\hphantom{0}	&	2005/08/28	&	{\it 4}\\
*Ivan	&   400	&	0.2\hphantom{0}\hphantom{0}	&	\hphantom{0}2.97	&	0.60	&	2004/09/14	&	{\it 4}\\
Model cyclone &	500	&	0.21\hphantom{0}	&	\hphantom{0}0.88	&	0.185	&	      	        &	{\it 8}\\
*Lili 	&700-900	&	0.52\hphantom{0}	&		&		&2002/09/15-27	      	        &	{\it 9}\\
*Isidore       &900-1300	&	0.47\hphantom{0}	&		&	 	& 2002/09/22-10/04	      	        &	{\it 9}\\
\bottomrule
\end{tabular}
\begin{tablenotes}
      \small
      \item References: {\it 1} -- \citet{miller62}; {\it 2} -- \citet{yang11}; {\it 3} -- \citet{braun06};
{\it 4} -- \citet{trenberth07}; {\it 5} -- \citet{fritz14};
{\it 6} -- \citet{riehl61}; {\it 7} -- \citet{huang2014}; {\it 8} -- \citet{kurihara75}; {\it 9} -- \citet{jiang08}. Estimates provided by \citet{hawkins68}, \citet{hawkins76}
and \citet{gamache93} (0.3-0.5 for $\overline{E}/\overline{P}$ for $r \lesssim 100$~km) are not listed as they are considered to be overestimates \citep{braun06}.
    \end{tablenotes}
  \end{threeparttable}
\end{table}

\section{Water vapor budget of a tropical cyclone}
\label{sec2}

\begin{figure*}[h!]
\centerline{\includegraphics[width=12.9cm]{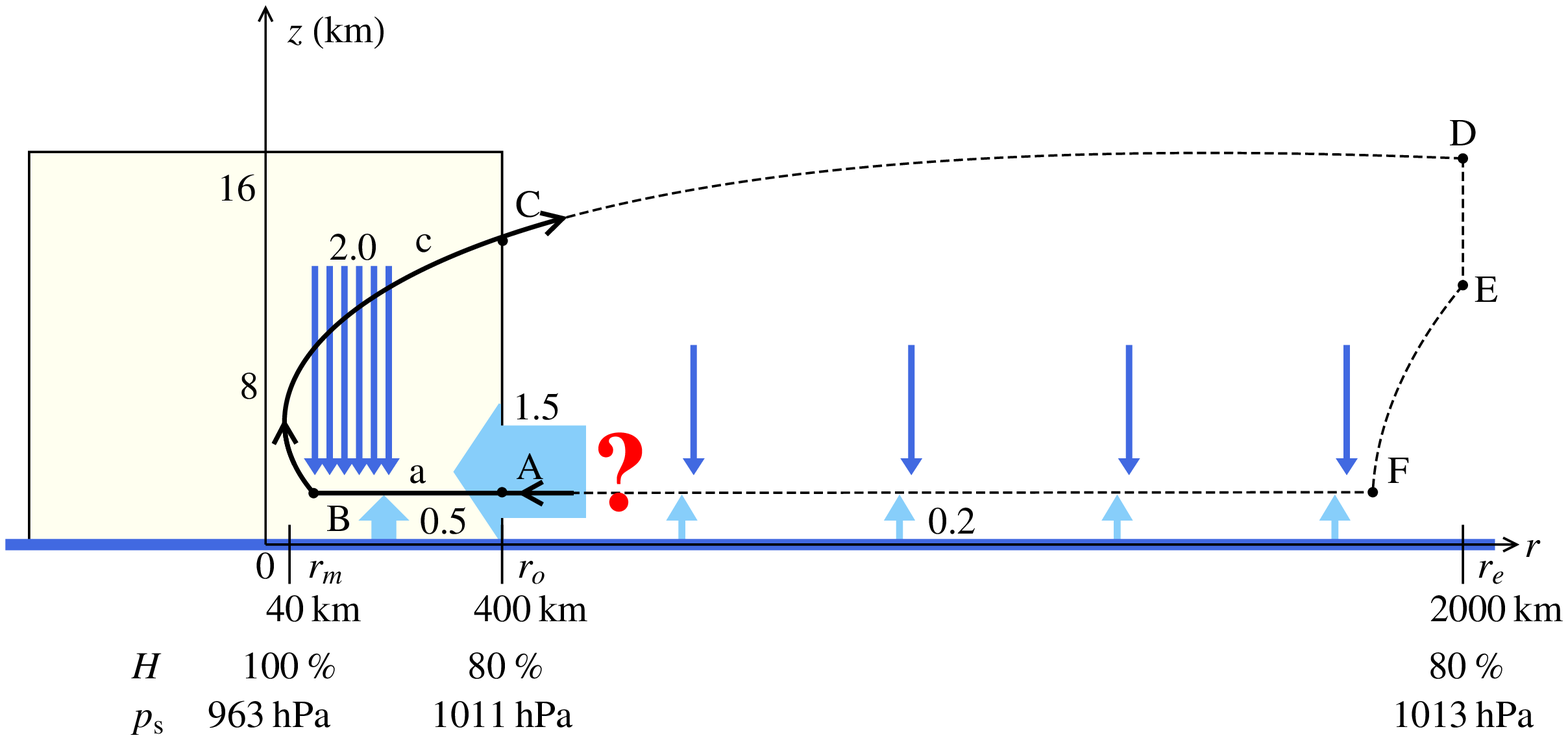}}
\caption{
A scheme for hurricane water vapor budgets based on Table~1 and the data for 1551 observations of North Atlantic hurricanes
in $1998$-$2015$ (see Methods): $r_m = 41\pm 24$~km
is the radius of the maximum wind velocity ($\pm$ one standard deviation); $r_o = 396\pm 121$~km is the radius of the outermost closed isobar;
$r_e \sim 2000$~km is a distant radius representing the mean tropical environment; $H$
and $p_s$ are relative humidity and air pressure at the surface $z = 0$; typical values for $H$ at
$r_m$, $r_o$ and $r_e$ are shown \citep{holland97} as well as $p_s(0) = 963 \pm 19$~hPa,
$p_s(r_o)= 1011 \pm 3$~hPa and $p_s(r_e)= 1013$~hPa as mean tropical sea level pressure.
The box represents the hurricane, black arrows represent air streamlines.
Values by upward blue arrows indicate typical mean evaporation (mm~h$^{-1}$) in the outer region $r \geqslant r_o$
and the inner region $r \leqslant r_o$ (1 mm~h$^{-1}$ corresponds to 1~kg~H$_2$O~m$^{-2}$~h$^{-1}$).
Horizontal blue arrow: typical import of water vapor into the hurricane (1.5~mm~h$^{-1}$ for $r \leqslant r_o$). Downward arrows: mean rainfall
within the inner area $r \leqslant r_o$ equals 2.0~mm~h$^{-1}$ (import plus evaporation); on larger areas mean rainfall approaches
mean evaporation (see text).
Dashed streamlines and the question mark at the horizontal blue arrow indicate that the origin of imported moisture remains unknown.
}
\label{hur}
\end{figure*}

We start by expressing evaporation-to-precipitation ratios via measurable atmospheric parameters.
Consider an air parcel containing $\tilde{N}$ moles of air,
$\tilde{N_d}$ moles of dry air and $\tilde{N_v}$ moles of water vapor, $\tilde{N} = \tilde{N}_d + \tilde{N}_v$.
It begins spiralling in the boundary layer from outside the hurricane
towards the area of maximum winds (the windwall), see points $\rm A$ and $\rm B$ in Fig.~\ref{hur}.
If our air parcel gathers water vapor evaporating from the oceanic surface beneath the parcel's trajectory,
the ratio $\gamma_d \equiv \tilde{N_v}/\tilde{N_d} = N_v/N_d$ must grow. Here $N_v \equiv \tilde{N_v}/\tilde{V}$ and
$N_d \equiv \tilde{N_d}/\tilde{V}$ (mol~m$^{-3}$) are molar densities of water vapor and dry air, respectively; $\tilde{V}$
is the parcel's volume.

Using the ideal gas law
\beq\label{ig}
p = NRT,
\eeq
where $p = p_v + p_d$ is air pressure, $p_v$ and $p_d$ are partial pressures of water vapor and dry air,
$T$ is temperature, $N = N_v + N_d$ is air molar density and $R = 8.3$~J~mol$^{-1}$~K$^{-1}$ is
the universal gas constant, the definitions of $\gamma_d \equiv N_v/N_d = p_v/p_d$, $\gamma \equiv p_v/p$ and
relative humidity $H \equiv p_v/p_v^*$, where $p_v^*$ is saturated pressure of water vapor,
we have:
\beq\label{dg}
\frac{d\gamma_d}{\gamma_d} = \frac{1}{1-\gamma} \left(\frac{dH}{H} - \frac{dp}{p} +\frac{dp^*_v}{p^*_v} \right).
\eeq
This relationship allows one to diagnose relative changes of $\gamma_d$ from the observed changes in pressure, temperature (which controls $p_v^*$)
and relative humidity. Since $\gamma \equiv p_v/p \sim 10^{-2} \ll 1$,
we can assume $\gamma = \gamma_d$ and neglect $\gamma$ in the right-hand side of Eq.~(\ref{dg}).

If, as is commonly assumed, the temperature of the air parcel does not change \citep[e.g.,][]{emanuel86}, we have
$dp^*_v/p_v^*=0$. Using typical values of $H$ and $p$ from Fig.~\ref{hur} we find
\beq\label{AB}
\frac{\Delta_{\rm AB}\gamma}{\gamma_{\rm B}} = \frac {H_{\rm B} - H_{\rm A}}{H_{\rm B}}  -\frac {p_{\rm B} - p_{\rm A}}{p_{\rm B}}= 0.25.
\eeq

As the air reaches the windwall, it ascends. If
the ascending air reaches a sufficient height, as occurs in intense cyclones, it
cools to such low temperatures that all water vapor condenses: $\gamma_{\rm B} \gg \gamma_{\rm C} \approx 0$ (Fig.~\ref{hur}).
Practically all this condensed moisture precipitates within the area $r \leqslant r_o$ \citep{jas13,huang2014,braun06}.
Thus, we have
\beq\label{BC}
\frac{\Delta_{\rm BC}\gamma}{\gamma_{\rm B}} = \frac{\gamma_{\rm C}}{\gamma_B} - 1 \approx -1.
\eeq
Thus, every air parcel entering the hurricane at point $\rm A$
and leaving it at point $\rm C$ (Fig.~\ref{hur}) acquires $\Delta_{\rm AB}\gamma$ moles
of water vapor by evaporation from the ocean and loses $-\Delta_{\rm BC}\gamma$ moles of water vapor
due to condensation and precipitation per each mole of dry air it contains. Therefore,
for the ratio of the mean evaporation $\overline{E}$ to mean precipitation $\overline{P}$
within area $r \leqslant r_{\rm C} = r_{\rm A}$ we have
\beq\label{rat}
\frac{\overline{E}}{\overline{P}} = -\frac{\Delta_{\rm AB}\gamma}{\Delta_{\rm BC}\gamma} = 0.25.
\eeq

In intense real-world cyclones the air temperature near the oceanic surface is not constant but may decline
towards the eyewall reflecting the rapid expansion of air parcels entering the area of lower pressure.
For example, in hurricane Isabel the surface air temperature drops by $-\Delta_{\rm AB} T =2$~K
as the air approaches the windwall from the periphery \citep{montgomery06}.
This corresponds to a negative third term in the right-hand part of Eq.~(\ref{dg}) of the order of
$\Delta_{\rm AB} p^*_v/p_v^* = (L/RT) \Delta_{\rm AB} T/T \sim -0.12$ for $T = 300$~K,
where $L = 45$~kJ~mol$^{-1}$ is the latent heat of vaporization.
In this case from Eq.~(\ref{dg}) we have $\Delta_{\rm AB}\gamma/\gamma_B = 0.2+0.05-0.12=0.13$. This means that 87\% of moisture
condensing within the hurricane must be imported from elsewhere.
As Eq.~(\ref{dg}) shows,
if the temperature drop is just slightly larger and the relative humidity increase
is slightly less than in our example above, the positive and negative terms in Eq.~(\ref{dg}) may cancel yielding
zero evaporation. In consequence, observational estimates of oceanic evaporation possess
large uncertainties (see, e.g., Table~1 of \cite{bell12}).

Since no direct measurements of surface water vapor flows in tropical cyclones exist \citep{trenberth03},
available estimates of oceanic evaporation either rely on the same governing relationship~(\ref{dg}) for various air streamlines
\citep[e.g.,][]{bell12} or consider the mass budget of the hurricane by comparing observed rainfall rates to estimates of imported
moisture (Table~1). Moisture convergence can be estimated from the experimentally measured or modelled wind field around the cyclone.
Using such estimates we can verify how Eq.~(\ref{dg}) explains the observed relationships.

\begin{figure*}[h!]
\centerline{\includegraphics[width=11cm]{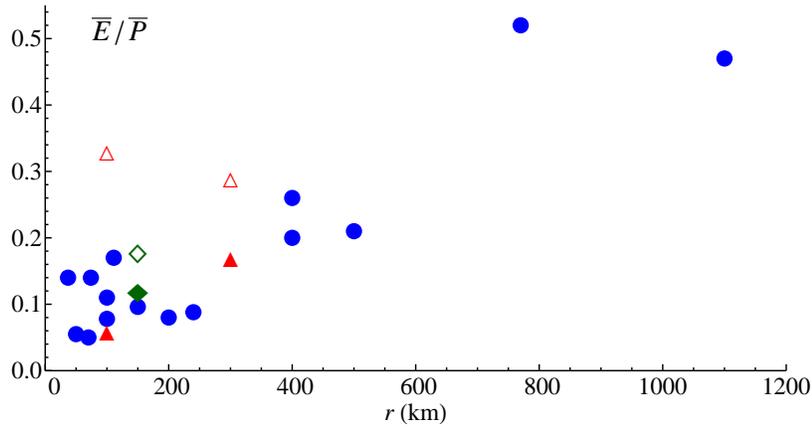}}
\caption{
Available estimates of evaporation-to-rainfall ratio for tropical cyclones as dependent on distance $r$ from
the cyclone center; $\overline{E}$ and $\overline{P}$ are the mean evaporation and precipitation
within a circle of radius $r$, respectively (Table~1). Empty and filled triangles: tropical storm Fay
on 1998/08/14 (lower rainfall) and 1998/08/15 (higher rainfall), respectively. Empty and filled diamond:
hurricane Daisy on 1958/08/27 (lower rainfall) and 1958/08/25 (higher rainfall), respectively.
Circles -- the remaining data from Table~1.
}
\label{evap}
\end{figure*}

First, Eq.~(\ref{dg}) predicts that in weak storms, where the air does not rise as high and
associated condensation in the rising air is incomplete, we have $-\Delta_{\rm BC} \gamma < \gamma_{\rm B}$,
and the evaporation-to-precipitation ratio (\ref{rat}) can grow. This pattern is reported for the tropical storm Fay \citep{fritz14}.
At the initial stage of the storm development, rainfall within 300~km was 0.7~mm~h$^{-1}$ and
the estimated evaporation-to-rainfall ratio was 0.33 and 0.29 within 100 and 300 km, respectively (Fig.~\ref{evap}).
As the storm intensified, rainfall within 300~km doubled, while $\overline{E}/\overline{P}$ declined
to 0.059 and 0.17 within 100 and 300~km, respectively (Table~1 and Fig.~\ref{evap}). Likewise, hurricane Daisy
at higher rainfall had a smaller $\overline{E}/\overline{P}$ than at lower (Table~1 and Fig.~\ref{evap}).

Second, Eq.~(\ref{dg}) explains why $\overline{E}/\overline{P}$ diminishes in the region
$r_m < r < r_o$ as compared to $r > r_o$(Fig.~\ref{evap}).
This is because for $r < r_o$ we only consider a certain part of the total pressure fall
and relative humidity increment, while still accounting for most of condensation and precipitation which take place in the ascending air in a narrow zone near
the windwall (consider path aBc in Fig.~\ref{hur}). In other words, $\Delta_{\rm aB} H$ and $-\Delta_{\rm aB} p$ are smaller than
$\Delta_{\rm AB} H$ and $-\Delta_{\rm AB}$ in Eq.~(\ref{AB}), while $\gamma_{\rm c} \approx \gamma_{\rm C}$ and $\Delta_{\rm Bc}\gamma \approx \Delta_{\rm BC} \gamma$.
Thus, $-\Delta_{\rm aB}\gamma /\Delta_{\rm Bc} \gamma < -\Delta_{\rm AB}\gamma /\Delta_{\rm BC} \gamma$, cf.~Eq.~(\ref{rat}).
Therefore, for $r < r_o$ we can observe small ratios $\overline{E}/\overline{P}< 0.25$. This agrees with the available
estimates (Table~1 and Fig.~\ref{evap}).

\begin{figure*}[h!]
\centerline{\includegraphics[width=12.9cm]{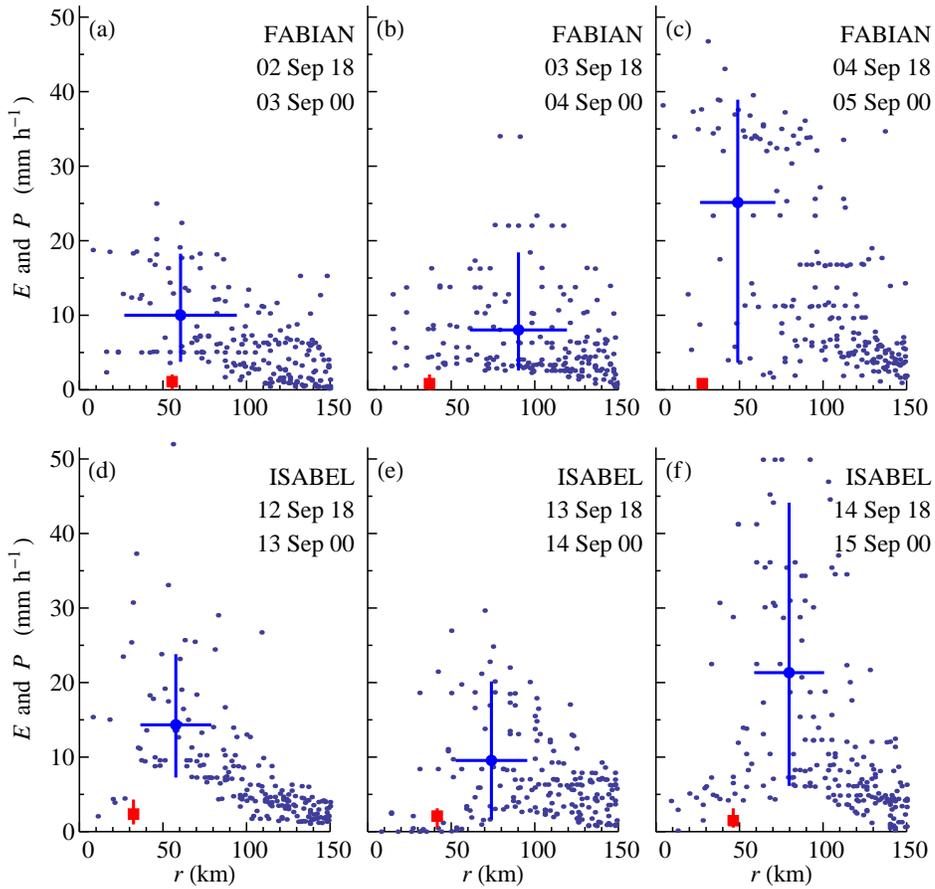}}
\caption{\label{isab}
Local precipitation $P$ and evaporation $E$ versus distance $r$ to the hurricane's center
in hurricanes Fabian (a-c) and Isabel (d-f) (see Methods for details). The dots show rainfall in individual grid cells
of the TRMM dataset. Date and time (UTC) are for the two EBTRK positions closest to the measurements
of \citet[][see their Fig.~5]{bell12}.
The big circles are the mean $P$ for $r_P - \Delta r_P \leqslant r \leqslant r_P + \Delta r_P$,
where $\Delta r_P$ is the standard deviation of radius $r_P$ of maximum precipitation.
The horizontal lines at the circles correspond to $r_P \pm \Delta r_P$,
the vertical lines connect the lower and upper 10\% quantiles of $P$.  The vertical
red lines show the full range of $E$ estimates from Fig.~15 of \citet{bell12} and the red squares are their mean values.
}
\end{figure*}

In the vicinity of hurricane's center -- in the hurricane's eye -- the air descends and rainfall is absent.
Thus, if evaporation is estimated from Eq.~(\ref{dg}) for $r < r_m$ (Fig.~\ref{hur}),
it can locally exceed rainfall.
This is illustrated by the available estimates of local oceanic evaporation in hurricanes
Fabian and Isabel in 2003 \citep{bell12}.
We calculated evaporation $E$ from the
data of Fig.~15 of \citet{bell12} on the oceanic enthalpy flux $Q_O$ (W~m$^{-2}$)
assuming that the evaporation latent heat flux accounts for three quarters of $Q_O$, $E L = 0.75 Q_O$ \citep{jaimes15}, and
using the conversion coefficient 1~mm~h$^{-1}$ = 696~W~m$^{-2}$ \citep{trenberth03}.
In these hurricanes the radius $r_P$ of maximum precipitation, which we defined
as the mean radius of the 10\% highest rainfall values, is about twice
the radius of maximum wind (Fig.~\ref{isab}). Thus, local estimates of evaporation can be close to or exceed local rainfall,
$E \approx P$ for $r < r_P$ (see, e.g., Fig.~\ref{isab}e), while the mean ratio $\overline{E}/\overline{P}$ within the area
$r < r_P$ can remain low.

Finally, Eqs.~(\ref{dg})-(\ref{rat}) do not provide support to the idea
that the moisture imported into the hurricane at $r < r_o$ derives from local evaporation in the region $r > r_o$. In this outer area
neither surface pressure nor relative humidity change appreciably with $r$ as compared to their changes
within the hurricane (Fig.~\ref{hur}). Thus, as the air parcel moves along the surface towards the hurricane center (consider path FA
in Fig.~\ref{hur}), it neither gathers nor loses water vapor, i.e. its $\gamma$ does not change. This means that such an air
parcel neither imports nor exports moisture to/from the oceanic area above which it is moving.

This means that for $r> r_o$, i.e. outside the area of considerable changes in surface pressure and humidity,
locally evaporated moisture precipitates locally. Obviously, with $r_o < r \to \infty$, the
evaporation-to-rainfall ratio should rise towards unity. Thus, for hurricanes Isidore and Lili the ratio assessed for $r \sim 1000$~km
is about $0.5$ (Table~1 and Fig.~\ref{evap}).

If not from evaporation that occurs over the ocean simultaneously with the hurricane development,
how does the hurricane get moisture to sustain its rainfall?

\section{Methods}
\label{met}

We used the "extended best track" dataset (EBTRK) released 27 July 2016 \citep{demuth06}. We picked up all observations
in $1998$-$2015$ of Atlantic tropical cyclones with maximum velocity $V_{\rm max} > 34$~m~s$^{-1}$ (hurricanes)
south of 40$^{\rm o}$~N, a total of 1551 observations for 122 different storms. Mean radii of maximum wind $r_m$
and outermost closed isobar $r_o$, mean sea level pressure in hurricane's center and
at $r=r_o$ were calculated from EBTRK data (Fig.~\ref{hur}). Translation velocity $V_c$,
i.e. velocity at which the cyclone center moves relative to the Earth's surface,
 was
calculated for each observation as the distance between the current and next position of the hurricane's center divided
by six hours (EBTRK data are recorded every six hours). The average translation velocity
was $V_c = (5.9 \pm 2.9)$~m~s$^{-1}$ ($\pm$ standard deviation).


We used 3-hourly TRMM 3B42 (version 7) and 1-hourly MERRA MAI1NXINT datasets to calculate
local rainfall $P(r)$ (mm~h$^{-1}$) and total water vapor in the atmospheric column $\sigma(r)$ (mm)
depending on distance $r$ from the hurricane's center.
Local values $P(r_i)$ and $\sigma(r_i)$ were defined as the mean rainfall and
water vapor content in all grid cells with $r_{i-1} \leqslant  r < r_{i}$, $1 \leqslant  i \leqslant  120$,
$r_i \equiv 25\times i$~km, $r_0 \equiv 0$~km, $r_{120} = 3000$~km.

Tropical rainfall and atmospheric moisture vary significantly in space and time. Hurricanes, too,
happen at different time of the year in different parts of the Atlantic ocean.
To distinguish the hurricanes' moisture footprint from regional, seasonal and diurnal variability
we introduced reference functions $X_0(r)$, $X = \{P, \sigma\}$.
They were calculated as the mean value of the considered variable for those seventeen years in
the eighteen years' $1998$-$2015$ period when the hurricane {\it did not happen}.

For each data entry we also calculated the same variables three days before and three days after the hurricane,
denoting them, respectively, $X^-$ and $X^+$ ($X = \{P, \sigma,P_0, \sigma_0 \}$).

For example, for hurricane Isabel observed 2010/09/13 at 18 UTC at $22.5^{\rm o}$N  $62.1^{\rm o}$W
(EBTRK data entry AL1303) we calculated the dependence of rainfall $P$ on
distance $r$ from this point on 13 September at 18 UTC in all years except $2010$. The mean of the obtained
seventeen $P(r)$ dependencies was defined as $P_0(r)$.
Variable $P^+_0(r)$ describes the dependence of rainfall on distance $r$ from the same point ($22.5^{\rm o}$N  $62.1^{\rm o}$W) on 16 September  at 18 UTC
(three days after the hurricane) averaged over all years, $1998$-$2015$, except $2010$.
Variable $\sigma^-(r)$ describes the dependence of local atmospheric
moisture content on distance $r$ from the same point ($22.5^{\rm o}$N  $62.1^{\rm o}$W) on 10 September 2010 at 18 UTC
(three days before the hurricane).

Finally, for all introduced variables $X = \{P, P^\pm, \sigma, \sigma^\pm, P_0, P_0^\pm, \sigma_0, \sigma_0^\pm \}$
we calculated $\overline{X}(r) \equiv 2 \pi \int_0^r X(r') r' dr' /\pi r^2$.
Everywhere below the overbar $\overline{X}(r)$ denotes the average of the considered variable $X(r)$ in the circle
of radius $r$. Variables without overbars, $X(r)$, correspond to mean values at a distance $r$ from the hurricane's center.
Subscript "$0$" denotes reference variables calculated in hurricane absence to describe an average hurricane-free environment.

Thus, for each of the 1551 analyzed EBTRK data entries we
obtained twenty four functions -- $P(r)$, $\overline{P}(r)$, $\sigma(r)$, $\overline{\sigma}(r)$,
$P_0(r)$, $\overline{P}_0(r)$, $\sigma_0(r)$, $\overline{\sigma}_0(r)$, $P^\pm(r)$, $\overline{P}^\pm(r)$, $\sigma^\pm(r)$, $\overline{\sigma}^\pm(r)$,
$P^\pm_0(r)$, $\overline{P}^\pm_0(r)$, $\sigma^\pm_0(r)$, $\overline{\sigma}^\pm_0(r)$. Their values averaged for each $r$ over all
analyzed EBTRK data entries are shown in Figs.~\ref{WV} and \ref{RR}.

The spatial resolution of the TRMM and MERRA datasets are $0.25^{\rm o}$ latitude $\times$ $0.25^{\rm o}$ longitude
and $0.5^{\rm o}$ latitude $\times$ $0.67^{\rm o}$ longitude, respectively.
Since rainfall is highly non-uniform spatially and temporally, rainfall observations in a single TRMM grid cell are
characterized by high uncertainty. For example, for most grid cells with zero rainfall rate the error is $0.35$~mm~h$^{-1}$; for
grid cells with a high rainfall rate the error is usually several times the rate itself. Thus, a reliable rainfall estimate demands at least a hundred
observations.

The standard error for a given $r$ of the rainfall variables averaged across all analyzed EBTRK data entries (Fig.~\ref{RR})
was calculated following \citet{huffman97} as $s_P(r) = \sqrt{\sum_{i=1}^{n(r)} s_i^2}/n(r)$.
Here $s_i$ is the error of the $i$-th rainfall value as recorded in the TRMM dataset
and $n(r)$ is the total number of TRMM rainfall values analyzed for a given $r$ for a given variable:
$n(r) = \sum_{j=1}^{m N_h} k_j$, where
$N_h = 1551$, $m = 17$ for variables corresponding to hurricane absence (17 years) and $m = 1$ otherwise,
and $k_j \geqslant 0$ is the number of rainfall values for a given $r$ for the $j$-th observation
(here "observation" is a unique combination  of coordinate, date and time).

Compared to rainfall, the atmospheric water vapor content is a much less variable quantity. We estimated the standard error
of the water vapor variables as $s_\sigma(r) = \beta(r) /\sqrt{N_h(r)}$, where $\beta(r)$ is the standard deviation of
$N_h(r)$ values of the considered variable for a given $r$; $N_h(r) \leqslant 1551$ is the number of
hurricane observations for each $r$.

For the water vapor variables for some hurricanes the data were
missing for $r\leqslant  75$~km, which is close to the spatial resolution of the MERRA dataset. The minimum number of data analyzed was
$N_h(r) = 817$ for $r = 25$~km, with each hurricane observation represented by one $\sigma(r)$ value. The number of data for a given $r$ for each
hurricane observation increases with growing $r$ and is larger for the spatially averaged variables $\overline{X}(r)$ compared
to local variables $X(r)$. For $r = 300$~km the mean number of values analyzed for each hurricane observation was $13\pm 3$
and $90 \pm 6$ for $\sigma(r)$ and $\overline{\sigma}(r)$
and $70\pm 5$ and $478\pm 30$ for $P(r)$ and $\overline{P}(r)$ ($\pm$ one standard deviation),
with totals of, respectively, $20471$, $138966$, $109279$ and $740759$. Our main results pertain to 300~km $< r < 1500$~km.

\section{Hurricane's long-range moisture signatures}
\label{sig}

\subsection{Hurricane's dry footprint}

\begin{figure*}[h!]
\centerline{\includegraphics[width=12.9cm]{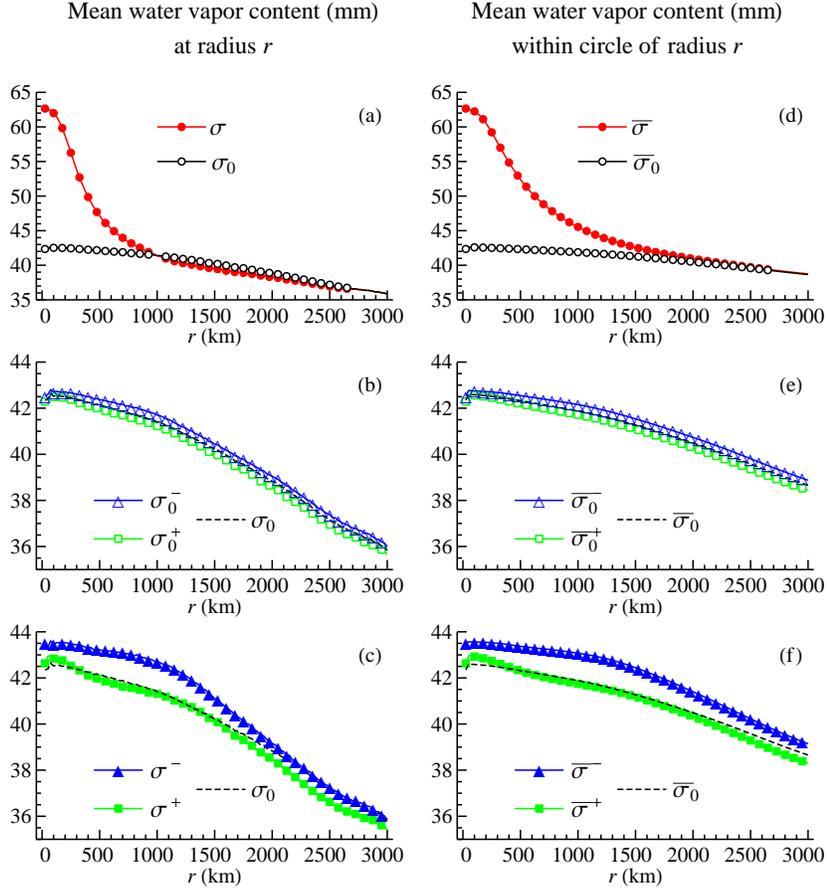}}
\caption{\label{WV}
Mean water vapor content versus radial distance $r$ from a hurricane's center
for 1551 observations of North Atlantic hurricanes in $1998$-$2015$ (see Methods for the description of variables).
(a)-(c) mean water vapor content at radius $r$; (d)-(f) mean water content within circle of radius $r$
(variables with overbars). Filled symbols and empty symbols (variables with subscript $0$) correspond to the presence
and absence of hurricanes, respectively. Superscripts "$-$" and "$+$" correspond to the water vapor distribution three days before and three days
after the hurricane, respectively. In each panel, the pair of solid curves lacks symbols at those $r$ where
the difference between the two variables is less than twice the sum of their standard errors (see Methods).
}
\end{figure*}

\begin{figure*}[h!]
\centerline{\includegraphics[width=12.9cm]{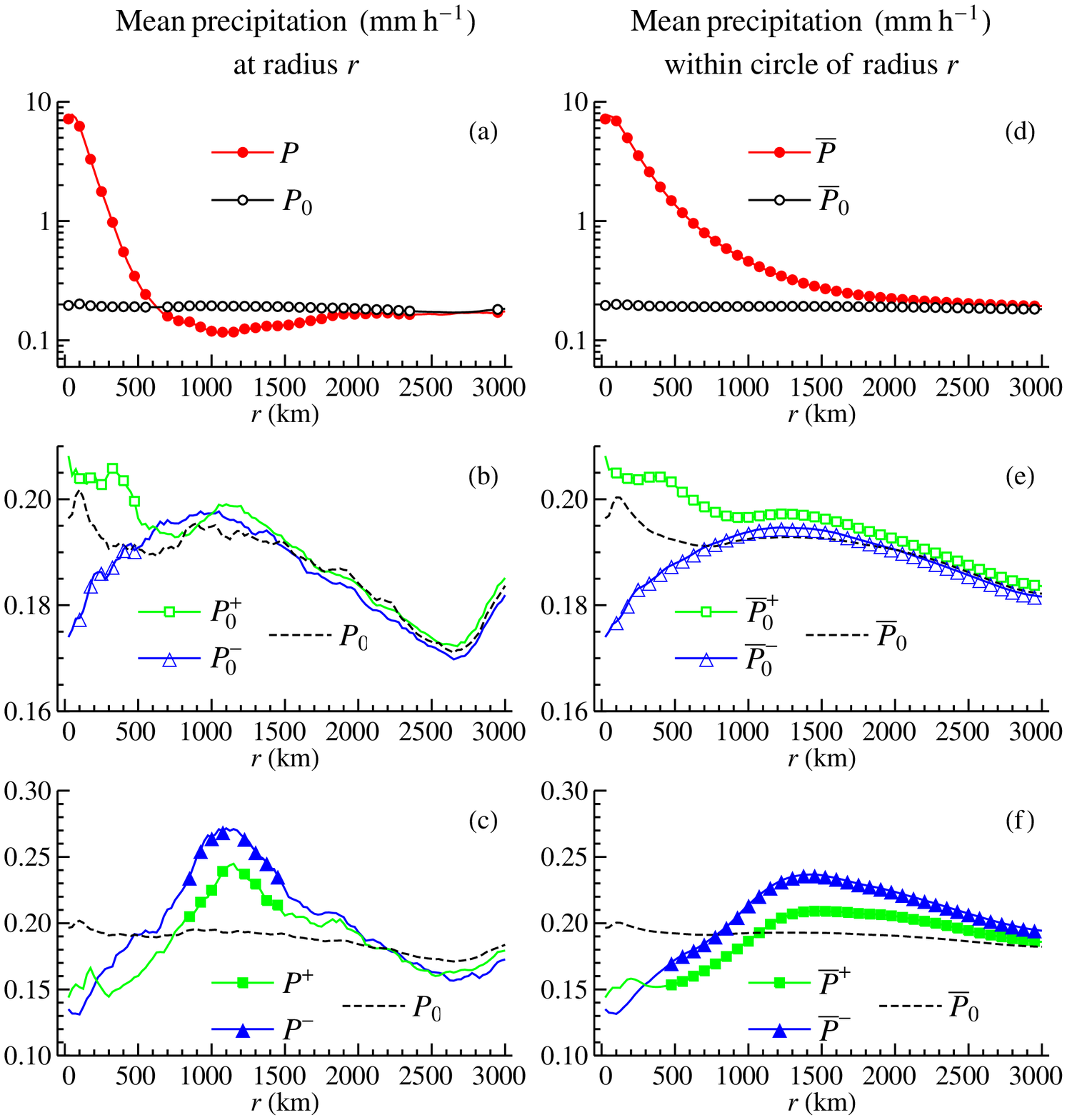}}
\caption{\label{RR}
Same as in Fig.~\ref{WV} but for rainfall rate. Note the logarithmic scale on the vertical axis in (a) and (d).
The two precipitation maxima in (c) are an indication of where the hurricane center is on average located --
relative to the current hurricane position -- three days before and three days after the hurricane.
}
\end{figure*}

We begin with discussing reference functions $\sigma_0(r)$ and $P_0(r)$ which characterize the hurricane region
in hurricane absence. In those years when hurricanes do not happen, the atmospheric water vapor content $\sigma_0(r)$
declines with increasing distance $r$ from the hypothetical hurricane center.
Its maximum and minimum values, $\sigma_0(r) = 42.6 \pm 0.006$~mm at $r = 75$~km and minimum $\sigma_0(r) = 35.8 \pm 0.001$~mm at $r = 3000$~km ($\pm$ one standard error),
differ by 19\% (Fig.~\ref{WV}a,d). This reflects the steeper temperature gradients and lower temperatures associated with large $r$ at higher latitudes.
A similar although less regular pattern is found for local reference rainfall: maximum $P_0(r)= 0.202\pm 0.005$~mm~h$^{-1}$ and minimum
$P_0(r)= 0.171\pm 0.001$~mm~h$^{-1}$ for $r = 100$~km and $r = 2650$~km, respectively, differ by 18\% (Fig.~\ref{RR}a,d).
In the absence of moisture import across $r = 3000$~km (the largest radius considered in this study),
the long-term mean spatially averaged rainfall $\overline{P}(r)$ for $r \leqslant 3000$~km equals the long-term mean
oceanic evaporation $E_0$ in the hurricane region, $E_0 = \overline{P}_0(3000~\rm km) = 0.182\pm 0.0001$~mm~h$^{-1}$ .

In the presence of hurricanes the water vapor content is higher, $\sigma(r) > \sigma_0(r)$ for $r < 10^3$~km.
This difference is due, at least in part, to the re-distribution of water vapor within the hurricane.
Because of the rapid vertical ascent of air near the eyewall $r \sim r_m$ (Fig.~\ref{hur}),
this region is saturated with water vapor, so $\sigma(r)$ rises for small $r$. At larger $r$, however, a slightly higher $\sigma(r)$
may reflect the fact that hurricane wind speeds {\it preferentially develop} when and where the amount of water vapor
is larger than usual. This is confirmed by the fact that water vapor content three days before the hurricane, $\sigma^-(r)$,
is on average higher than it is three days after the hurricane and than it is in hurricane absence, i.e. $\sigma^-(r)>\sigma^+(r)\approx \sigma_0(r)$
(Fig.~\ref{WV}c,f). The maximum difference $\sigma^-(r) - \sigma^+(r) = 1.5$~mm is reached for $r = 700$~km, where
$\sigma^-(r) = 43.1 \pm 0.04$~mm, $\sigma^+(r) = 41.6 \pm 0.04$~mm and $\sigma_0(r) = 41.9 \pm 0.03$~mm (Fig.~\ref{WV}c).
During the hurricane the moisture content at this radius rises to $\sigma(r) = 44.0 \pm 0.03$~mm (Fig.~\ref{WV}a).

Dry air expelled from the hurricane should descend somewhere thus suppressing rainfall and diminishing moisture content outside the hurricane.
Indeed, we find that in the presence of hurricanes, local rainfall $P(r)$ displays a remarkable behavior:
it is higher than $P_0(r)$ for $r < r_d = 625$~km and lower than $P_0(r)$ for $r \geqslant  r_d$, Fig.~\ref{RR}a.
(A qualitatively similar behavior, albeit at larger radii, is displayed by local water vapor content in the presence of hurricanes
$\sigma(r)$, Fig.~\ref{WV}a.)
The minimum is reached at $r = 1050$~km;
here $P(r) = 0.12\pm 0.004$~mm~h$^{-1}$ is only 60\% of $P_0(r)= 0.19\pm 0.001$~mm~h$^{-1}$.

The existence of this minimum is supported by the fact that local rainfall in the hurricane
center three days before and three days after the hurricane, $P^-(r)$ and $P^+(r)$, declines considerably
below the long-term mean $P_0(r)$, Fig.~\ref{RR}c. A hurricane moving with translation velocity $V_c = 6$~m~s$^{-1}$
covers about 1.5 thousand kilometers in three days.
Thus, according to Fig.~\ref{RR}a, the point which in three days from now becomes a new hurricane center is currently located within
the region of minimal rainfall of the approaching hurricane (see Fig.~\ref{RR}c). Therefore, rainfall in the considered point
three days before the hurricane is lower than the long-term mean, $P^-(0) < P_0(0)$.
(Indeed, \citet{anthes74} noted that "one of the earliest
recognized signs of an approaching storm is the unusual suppression of normal tropical cloudiness".
Figure~\ref{RR}c quantifies this effect.)
For the same reason the rainfall three days after the hurricane declines as well, $P^+(0) < P_0(0)$, Fig.~\ref{RR}c.

\begin{figure*}[h!]
\centerline{\includegraphics[width=12.9cm]{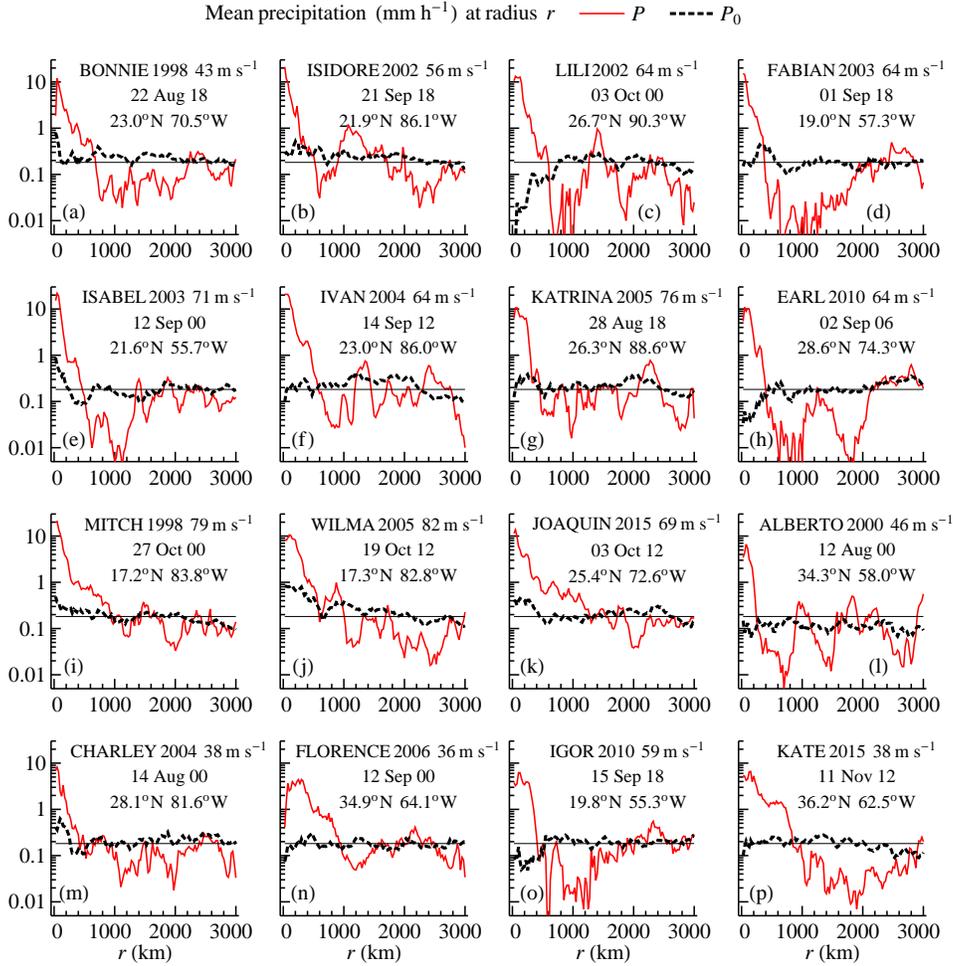}}
\caption{\label{Hurri}
Precipitation versus distance from the hurricane center in some North Atlantic hurricanes: (a)-(h) eight hurricanes discussed in
Section~2: Bonnie, Ivan, Katrina, Isidore, Lili (Table~1), Fabian, Isabel (Fig.~\ref{isab}), Earl \citep{jaimes15};
(i)-(k) three most intense hurricanes
in $1998$-$2015$ (one most intense hurricane in each six years is chosen),
(m)-(p) $(n\times310)$th ($n=1,2,3,4,5$)
from our set of $5 \times 310 + 1 = 1551$ analyzed observations (ordered by date) as examples
of randomly selected hurricanes. As in Fig.~\ref{RR}a, $P(r)$ is the actual rainfall distribution in the hurricane,
$P_0(r)$ is the rainfall distribution in those seventeen years when the hurricane did not occur
(see Methods). Horizontal line denotes $E_0 = \overline{P}_0(3000~\rm km) = 0.18$~mm~h$^{-1}$ from Fig.~\ref{RR}d.
Maximum wind velocity, the date and time of the observation and the coordinate of the hurricane center $r = 0$
are from the EBTRK dataset.
}
\end{figure*}

This rainfall minimum for $r > r_d$ is not just a property of the
statistical mean precipitation distribution averaged over all hurricane observations (Fig.~\ref{RR}a). In every hurricane
there is a region $r > r_d$ where rainfall $P(r)$ drops to values below the long-term mean rainfall $P_0(r)$
and long-term mean evaporation $E_0$,
although the magnitude of this drop and its precise location vary from hurricane to hurricane (Fig.~\ref{Hurri}).

For each hurricane observation, we defined $r_d \geqslant r_m$ as the minimal radius where $P(r_d)= E_0 = 0.18$~mm~h$^{-1}$
($r_m$ is the radius of maximum wind, Fig.~\ref{hur}). For $1545$ observations where $r_m$ is known in the EBTRK dataset,
we obtained $r_d = 502 \pm 230$~km ($\pm$ one standard deviation). This is lower than $r_d = 625$~km
estimated from the average rainfall distribution in Fig.~\ref{RR}a.
We can call $r_d \approx 500$~km the radius of the "dry footprint" left by the hurricane
in its wake: within the circle $r \leqslant r_d$ hurricane rainfall $P(r)$ rises above the long-term mean tropical
oceanic evaporation.

\begin{figure*}[h!]
\centerline{\includegraphics[width=9cm]{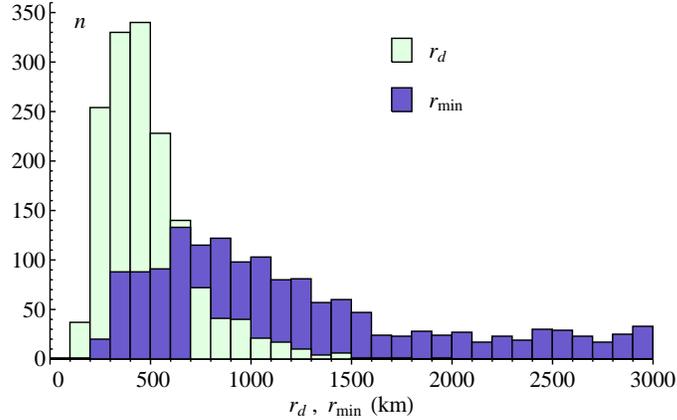}}
\caption{\label{rmin}
Frequency distribution of $r_d \geqslant r_m$, for which $P(r_d) = E_0 = 0.18$~mm~h$^{-1}$, and $r_{\rm min} \geqslant r_m$,
for which $P(r_{\rm min})$ is minimal, for $1545$ hurricane observations.
}
\end{figure*}

The mean radius where minimum rainfall is observed for
each hurricane observation for $r \geqslant r_m$ was $r_{\rm min} = 1206 \pm 725$~km, which is larger than $r_{\rm min} = 1050$~km
estimated from the average rainfall distribution in Fig.~\ref{RR}a.
The mean minimal rainfall was $P_{\rm min}(r_{\rm min}) = 0.011 \pm 0.15$~mm~h$^{-1}$, in good agreement
with the above discussed minimum of $P(r)$ in Fig.~\ref{RR}a.
The frequency distributions of $r_d$ and $r_{\rm min}$ show that
in most hurricanes local rainfall falls below the long-term mean evaporation between $300~{\rm km} < r \leqslant 500$~km,
while the minimum rainfall is observed within $600~{\rm km} < r \leqslant 900$~km (Fig.~\ref{rmin}).

\subsection{Rainfall and evaporation}
\label{r&e}

In contrast to local rainfall $P(r)$, which is either lower or higher than $P_0(r)$ depending on $r$,
mean rainfall $\overline{P}(r)$ within a circle of radius $r$ is always higher in the presence of hurricanes
than in their absence, $\overline{P}(r) \geqslant \overline{P}_0(r)$ for $r \geqslant 0$~km (Fig.~\ref{RR}d).
This has important implications for understanding a hurricane's water vapor budget.

Previous studies suggested that a hurricane can be sustained by evaporation on an area of a three-five times
larger radius than that of hurricane's rainfall area \citep{anthes74,trenberth03}. In the framework of our results, the rationale behind
such a statement can be formulated as follows. Within a hurricane of radius $r_o = 400$~km mean rainfall rate
$\overline{P}(r_o)=1.9$~mm~h$^{-1} = 1.9~\rm{kg~H_2O~~m^{-2}~h^{-1}}$ is over ten times the long-term mean evaporation $E_0 = 0.18$~mm~h$^{-1}$.
Total hurricane rainfall (kg~H$_2$O h$^{-1}$) equals $\Pi \equiv \pi r_o^2 \overline{P}(r_o)$. This rainfall
can be generated by the steady-state mean evaporation within an area $r \leqslant r_e$, where
$r_e$ is determined by the equality $\Pi = \pi r_e^2 E_0$.
This gives $r_e = r_o \sqrt{\overline{P}(r_o)/E_0} = 1300$~km.

However, this mass-balance consideration presumes that all water vapor produced by evaporation
on a larger area $r \leqslant r_e$ is concentrated to precipitate within the inner circle $r \leqslant r_o < r_e$.
Then there is no rainfall in the ring $r_o \leqslant r \leqslant r_e$. In reality
we have seen that mean local rainfall $P(r)$, while reduced at larger radii, never falls
below 60\% of $E_0$ (Fig.~\ref{RR}a). According to our analysis, mean local rainfall for $400~{\rm km} \leqslant r \leqslant 3000$~km is $0.89 E_0$.
This means that only $11\%$ of water vapor evaporated outside the hurricane could have
been imported into the inner circle $r \leqslant r_o$. Thus, the necessary radius which could ensure
the observed rainfall distribution within $r \leqslant r_o$ rises up to $r_e \approx r_o \sqrt{\overline{P}(r_o)/E_0/0.11} \approx 3900$~km.
There is no evidence that concentration of moisture at this scale -- an
area of nearly eight thousand kilometers in diameter -- occurs.

We also note that the observed suppression of rainfall for $r \geqslant r_d$ does not necessarily imply that a certain
part of evaporated moisture is exported from this area to the inner hurricane area. Rather, this is an indication of dry air masses
expelled from the hurricane descending outside and mixing with local air.
This would suppress local rainfall
and allow water vapor to accumulate in the atmosphere after its depletion by the cyclone.
\citet{frank77}, based on observations for northwest Pacific cyclones,
indicated that around $440$-$660$~km from the cyclone center there is a region of air subsidence throughout most of the troposphere. This radius
is close to radius $r_d$ of the hurricane's dry footprint that we estimated at $r_d \approx 500$~km in the previous section.
It corresponds to the so-called {\it moat} of clear skies that typically surround tropical cyclones.
It is at this radius that the water vapor drawn into the cyclone at the mid-level descends to the boundary layer to
be driven towards the eyewall by the boundary layer inflow and then condense in the air updrafts \citep{frank77}.

For $r = 1300$, $2000$ and $3000$~km we find that $\overline{P}(r)$ is, respectively,
$67$\%, $17$\% and 6\% higher than the long-term evaporation $E_0$.
Since long-term steady-state evaporation is limited by the available solar energy, these findings suggest
that a hurricane is not a steady-state system in the sense that,
while maintaining a high moisture content within itself (Fig.~\ref{WV}a),
 it depletes atmospheric moisture faster than it is replenished by evaporation on a scale of a few thousand kilometers.
This is supported by the observation that moisture content prior to hurricane formation is appreciably higher
than after the hurricane sweeps the area, $\sigma^-(r) > \sigma^+(r)$ (Fig.~\ref{WV}c,f).

We will now propose a distinct concept regarding how atmospheric moisture sustains a hurricane.  In this concept
there is no need to draw in moisture over thousands of kilometers but rain is sustained from atmospheric
stocks within about $700$~km of the storm's center.

\section{Water vapor budget and hurricane movement}
\label{mot}

\subsection{Governing relationships}
\label{sPm}

\begin{figure*}[h!]
\centerline{\includegraphics[width=10cm]{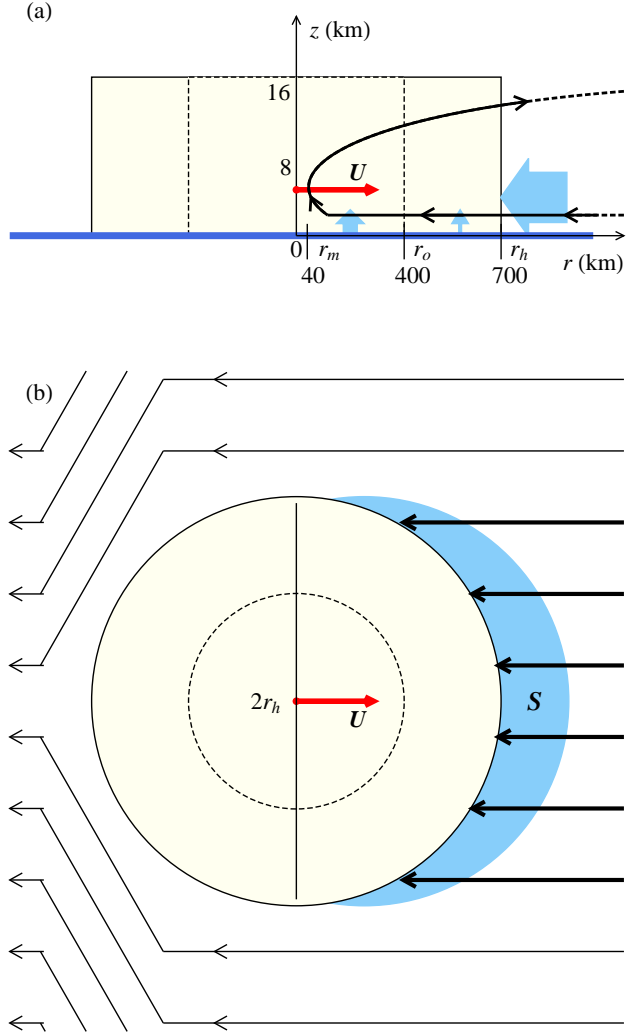}}
\caption{
(a) Water vapor budget in the hurricane rest frame (cf. Fig.~\ref{hur}). As the hurricane moves through
the atmosphere with velocity $U$ relative to the ambient air, it consumes the ambient water vapor. Dry air leaving the hurricane
in the upper atmosphere does not return.
(b) A bird's-eye view of (a) from the top of the inflow layer: $S = 2 r_h U \Delta t$
is the area swept by the hurricane's diameter $2r_h$ in time $\Delta t$. Water vapor
in the inflow layer on area $S$ is drawn into the hurricane, condenses and precipitates during the time interval $\Delta t$.
Thick air streamlines are those that bring in moisture and disappear within the hurricane
(the air ascends and leaves the hurricane above the inflow layer). Thin streamlines show air flowing around the
hurricane without entering it.
}
\label{hur2}
\end{figure*}

Can a hurricane sustain itself through its motion?
A hurricane of diameter $2r$ moving through the atmosphere
with velocity $U$ sweeps an area of $S = 2 r U\Delta t$ during a time interval $\Delta t$ (Fig.~\ref{hur2}).
The total amount of water vapor in this area in a hurricane-free environment is $2 r U\Delta t \sigma_0(r)$.
Let all this moisture condense and precipitate
within the hurricane area $\pi r^2$ during same time interval. Mean rainfall
in the circle of radius $r$ (mm~h$^{-1}$) will be
\beq\label{Pm}
\overline{P}_m(r) = \frac{2 r U\Delta t\sigma_0(r)}{\pi r^2 \Delta t} = \frac{U}{\pi} \frac{2}{r}\sigma_0(r).
\eeq
In the frame of reference co-moving with the storm, the import of water
vapor into the hurricane corresponds to mean radial velocity $V_r = U/\pi$:
water vapor flux captured by the hurricane's diameter $2 r$ is averaged over
circumference of length $2 \pi r$. Water vapor leaves the hurricane via rainfall over area $\pi r^2$.
So factor $2/r$ in Eq.~(\ref{Pm}) is the consequence of the continuity equation (matter conservation).

Note that $\overline{P}_m(r)$ is the maximum rainfall for a given $r$ available to the hurricane through its motion. This upper limit is reached if
all water vapor encountered by the hurricane over a distance of $2r$ is used up (condensed and precipitated) within it. In other words,
a non-zero velocity $U > 0$ is a necessary but insufficient condition for a hurricane to sustain its rainfall by motion.
If the air flows through the hurricane without losing moisture or if the air blows around the hurricane without
entering it, the motion-related rainfall can be zero however large $U$. In such a case $U > 0$ but $V_r = 0$ (see thin air streamlines in Fig.~\ref{hur2}b).

\begin{figure*}[h!]
\centerline{\includegraphics[width=9cm]{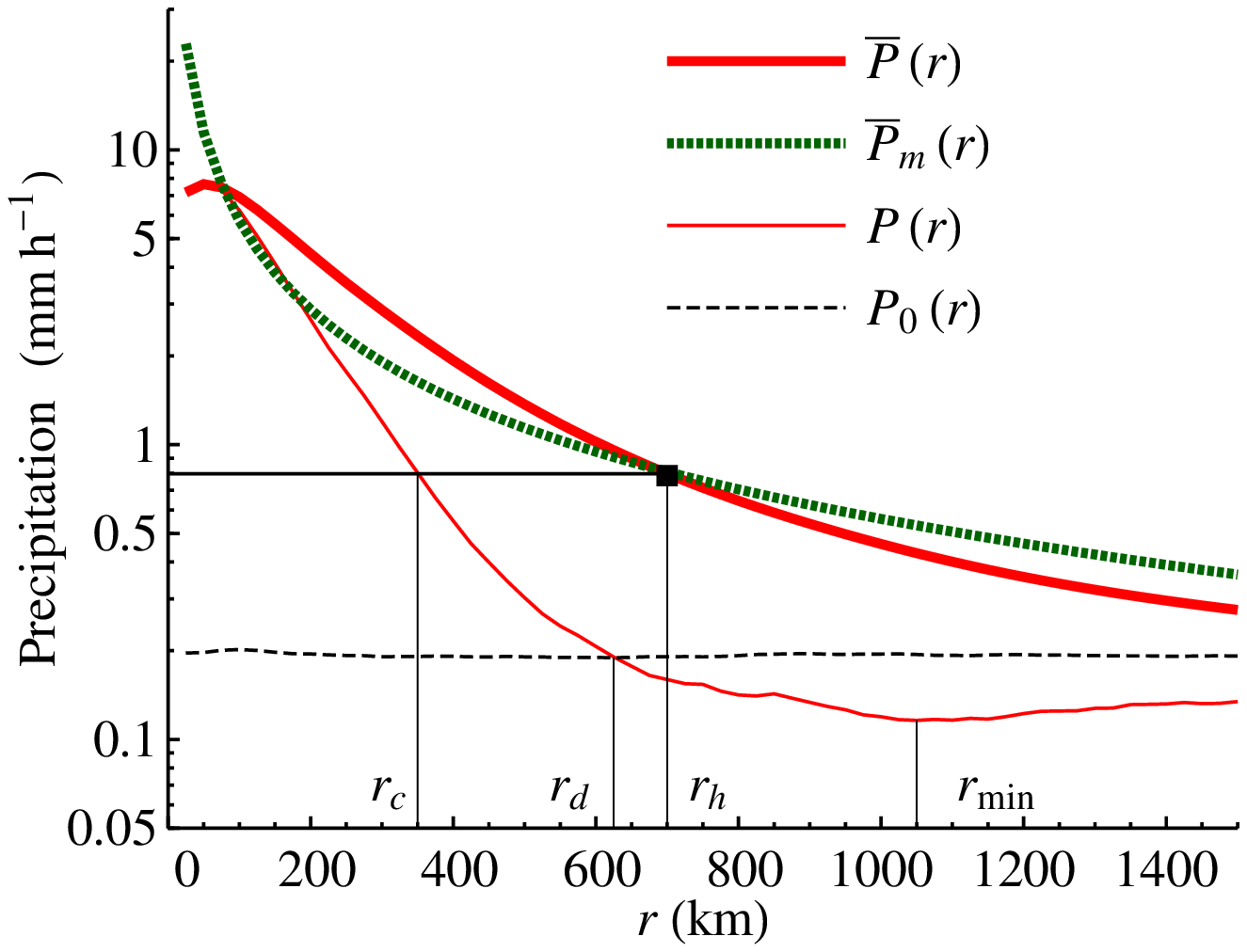}}
\caption{
Motion-related rainfall $\overline{P}_m(r)$ versus observed rainfall $\overline{P}(r)$ for $U = V_c$ in Eq.~(\ref{Pm}). Local rainfall $P(r)$ and
reference rainfall $P_0(r)$ from Fig.~\ref{RR}a are also shown to illustrate the relationship between
the inner radius $r_d$ of the hurricane's "dry footprint", for which $P(r_d) = P_0(r_d)$, the self-sufficiency radius $r_h$,
for which $\overline{P}(r_h) = \overline{P}_m(r_h)$, and the characteristic radius $r_c$
of water vapor re-distribution within the hurricane, for which $P(r_c) = \overline{P}(r_h)$ (see text for details).
}
\label{figPm}
\end{figure*}

Putting $U = V_c$ in Eq.~(\ref{Pm}), where $V_c = 5.9$~m~s$^{-1}$ is the mean translation velocity of our hurricanes,
and using $\sigma_0(r)$ from Fig.~\ref{WV}a in Eq.~(\ref{Pm}), we find that
motion-related rainfall $\overline{P}_m(r)$ equals observed rainfall $\overline{P}(r)$ for $r = r_h = 700$~km (Fig.~\ref{figPm}).
This means that water vapor available in an average hurricane-free environment
is sufficient to generate the observed hurricane rainfall within a circle of $r_h = 700$~km radius
{\it even if evaporation for $r \leqslant r_h$ is zero}. We can say that radius $r_h$ is the radius of a hurricane's self-sufficiency
with respect to water vapor: as long as the hurricane moves through the atmosphere with velocity $U$ equal to translation velocity
and consumes all ambient vapor it encounters, its rainfall is sustained. Such a hurricane does not need to concentrate moisture from afar.

For $100~\rm{km} \geqslant r < r_h$~km the observed rainfall $\overline{P}(r)$ is higher than the motion-related rainfall $\overline{P}_m(r)$
by up to $55$\% (Fig.~\ref{figPm}).
This reflects the radial concentration of the incoming moisture within the moving hurricane.
The radial velocity $V_r$ that delivers
water vapor to the inner area $r < r_h$ may increase with diminishing $r$ from its mean value of $V_r = U/\pi$
at $r = r_h$ to over twenty meters per second at its maximum in the vicinity of the windwall \citep{trenberth03,montgomery06,pla11}.
This concentration causes local rainfall $P(r)$ at $r < r_c$ to rise above the mean rainfall $\overline{P}(r_h)$,
while at larger radii it remains lower. Radius $r_c = 375$~km, where $P(r_c) = \overline{P}(r_h)$, appears to be
close to the radius of the outermost closed isobar, $r_c \approx r_o = 400$~km (Fig.~\ref{figPm}).
The ring $r_c = r_o \leqslant  r \leqslant  r_h$ serves as the donor of water vapor for the inner circle $r \leqslant r_c = r_o$.
The spatial scales linking water budgets and pressure/velocity distributions merit further investigations.

Real hurricanes would move through the atmosphere and obtain water vapor by motion more slowly, $U < V_c$.
On the other hand, evaporation within $r \leqslant r_h$, however small, is not zero. These factors,
as we show below, compensate each other such that the basic scenario we have just considered
turns out to produce a realistic self-sufficiency radius $r_h \approx 700$~km.

Expressing $U$ in terms of translation velocity as $U \equiv k_U V_c$, we define {\it available rainfall} $\overline{P}_a(r)$ within a circle of
radius $r$ as the sum of rainfall due to motion and rainfall due to evaporation that is concurrent with the hurricane:
\begin{equation}\label{Pa}
\overline{P}_a(r) = k_{\sigma} \overline{P}_m + k_E E_0 = k \frac{V_c}{\pi} \frac{2}{r}\sigma_0(r) + k_E E_0,
\end{equation}
where $k \equiv k_\sigma k_U$. Coefficient $k_\sigma \leqslant 1$ in Eq.~(\ref{Pa}) describes how completely
the hurricane uses up the store $\sigma_0(r)$ of atmospheric water vapor that it encounters.
Long-term evaporation $E_0=0.18$~mm~h$^{-1}$ does not depend on hurricane motion.
Coefficient $k_E(r) \geqslant 0$ describes how evaporation in hurricane presence varies relative to
the long-term mean; $\overline{E}(r) = k_E(r) E_0$ is the mean evaporation within the circle of radius $r$.

Before comparing available rainfall to observations we discuss
$k_U$, $k_\sigma$ and $k_E$ in Eq.~(\ref{Pa}).

\subsection{Key parameters}

\subsubsection{Thickness of the inflow layer}

\begin{figure*}[h!]
\centerline{\includegraphics[width=12.9cm]{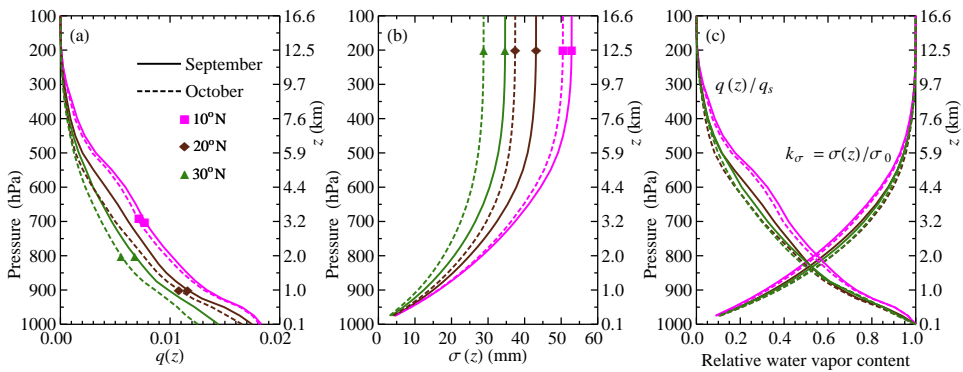}}
\caption{
Zonally averaged water vapor content as derived from the monthly MERRA dataset MAIMCPASM for $2015$:
Absolute values of $q(z)$ (a) and $\sigma(z)$ (b), see Eq.~(\ref{ks}),
in September and October (the months accounting for two thirds of all hurricane observations in
our analysis) at $10^{\rm o}$, $20^{\rm o}$ and $30^{\rm o}$~N;
(c) same functions as in (a) and (b) but each divided by its respective value
at $1000$~hPa for $q(z)$ (subscript $s$) and at $0.1$~hPa for $\sigma(z)$ (subscript $0$).
}
\label{WVH}
\end{figure*}

The value of $k_\sigma\leqslant 1$ describes the effective thickness of the inflow layer, where the hurricane
takes in the ambient water vapor (Fig.~\ref{hur2}a):
\beq
\label{ks}
k_\sigma(z) \equiv \frac{\sigma(z,r)}{\sigma_0(r)},\,\,\,\sigma(z,r)\equiv \int_0^z q(z',r) \rho(z',r) dz'.
\eeq
Here $q = \rho_v/\rho$ is absolute humidity (mass fraction of water vapor) (Fig.~\ref{WVH}a),
$\rho$ is air density and $\rho_v$ is water vapor density, $\sigma(z)$ is the amount of water vapor
vapor below altitude $z$ (Fig.~\ref{WVH}b) and $\sigma_0$ is the total amount of water vapor in the atmospheric
column. Since air temperature declines with height, the vertical distribution of water vapor is bottom-heavy: the warmer lower
7~km of the tropical atmosphere contain over 95\% of total water vapor in the atmospheric column (Fig.~\ref{WVH}c).
We have $k_\sigma(z) = 1$ if the inflow layer of height $z$ harbors all water vapor in the atmospheric column.

\begin{figure*}[h!]
\centerline{\includegraphics[width=12.9cm]{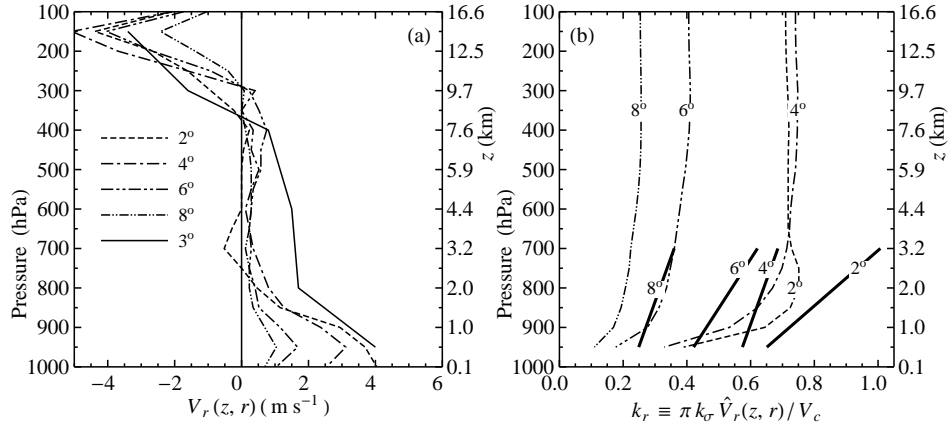}}
\caption{
Radial velocity $V_r$ versus height $z$ at selected radial distances $r$ (shown in degrees latitude).
(a) data for 2$^{\rm o}$, 4$^{\rm o}$, 6$^{\rm o}$ and 8$^{\rm o}$ are from Fig.~12 of \citet{frank77} for Northwest Pacific tropical cyclones;
data for 3$^{\rm o}$ -- from Fig.~9 of \citet{miller58} for North Atlantic hurricanes; (b) $k_r \equiv \pi k_\sigma \hat{V}_r(z,r)/V_c$
(see Eqs.~\ref{Vr} and \ref{kr}) for $V_c = 5.7$~m~s$^{-1}$; dashed curves -- data of \citet{frank77} from (c); solid curves -- data from Fig.~8 of \citet{miller58}
for Atlantic hurricanes.
}
\label{figVr}
\end{figure*}

The inflow into the cyclone in the region $r > r_o = 400$~km occurs up to the level of $400$-$500$~hPa or about $7$~km --
in contrast to the smaller radii where the inflow is confined to the lowest one and a half kilometer \citep{miller58,frank77,montgomery06}.
At a given $r$ this inflow appears to be stronger in hurricanes than it is on average in tropical cyclones (Fig.~\ref{figVr}a, solid line).
\citet{frank77} emphasized the importance of this mid-level inflow for the energy budget of tropical cyclones.
Most of this inflow should subside to the boundary layer "before being cycled upwards in clouds" at smaller radii \citep{frank77}.

The outflow in intense cyclones occurs above 10~km (Fig.~\ref{figVr}a).
The air expelled by the hurricane is nearly completely dry,
because at these heights water vapor content diminishes to negligible magnitudes of the order of $\gamma \sim 10^{-5}$ (Fig.~\ref{WVH}a).
This corresponds to $\gamma_{\rm C} = 0$ in Fig.~\ref{hur}.
Thus, a cyclone where the inflow occurs in the lower 7~km and outflow occurs above 10~km can suck in and use up practically all
atmospheric water vapor it encounters, i.e. for it $k_\sigma \approx 1$ in Eq.~(\ref{Pa}).
If the inflow occurs in the lower three kilometers, then $k_\sigma \approx 0.7$ (Fig.~\ref{WVH}c).

The ratio of $\sigma(z)/\sigma_0$ is practically independent of $\sigma_0$ (cf. Fig.~\ref{WVH}b and \ref{WVH}c). This justifies our
consideration of $k_\sigma$ (\ref{ks}) as a function of $z$ only.

\subsubsection{Propagation velocity $U$}

Velocity $U$ in Eqs.~(\ref{Pm}) and (\ref{Pa}) would be equal to the translation velocity $V_c$ of the cyclone, i.e. to the velocity
at which the cyclone center moves relative to the Earth's surface, if the storm moved through an otherwise still atmosphere.
When the hurricane is embedded into an air flow, velocity $U$ is equal to the difference between translation velocity $V_c$
and mean velocity $V_e$ of environmental flow. This velocity is sometimes referred to as {\it propagation velocity} \citep{carr90,franklin96}.
Propagation velocity $\mathbf{U}(r)$ at a given radius is equal to  translation velocity $\mathbf{V}_c$
minus mean velocity $\mathbf{V}_e(r)$ of air contained within a narrow ring $r,r+dr$: $\mathbf{U}(r) = \mathbf{V}_c -
\mathbf{V}_e(r)$.

Propagation velocity has been studied in the context of the {\it steering flow} concept, whereby the tropical
cyclone motion is considered as driven by the surrounding air flow \citep{george76,chan82,holland84,dong86,carr90,franklin96,
chan05,yasunaga16}. Propagation velocity is non-zero
when the cyclone deviates from the steering flow. Previous studies focused on finding
a horizontal scale and a pressure level where this deviation is minimized -- such that translation velocity $\mathbf{V}_c$
(and thus cyclone track) could be predicted from velocity $\mathbf{V}_e$ of the ambient air \citep{george76,holland84,franklin96}.

\begin{figure*}[h!]
\centerline{\includegraphics[width=12.9cm]{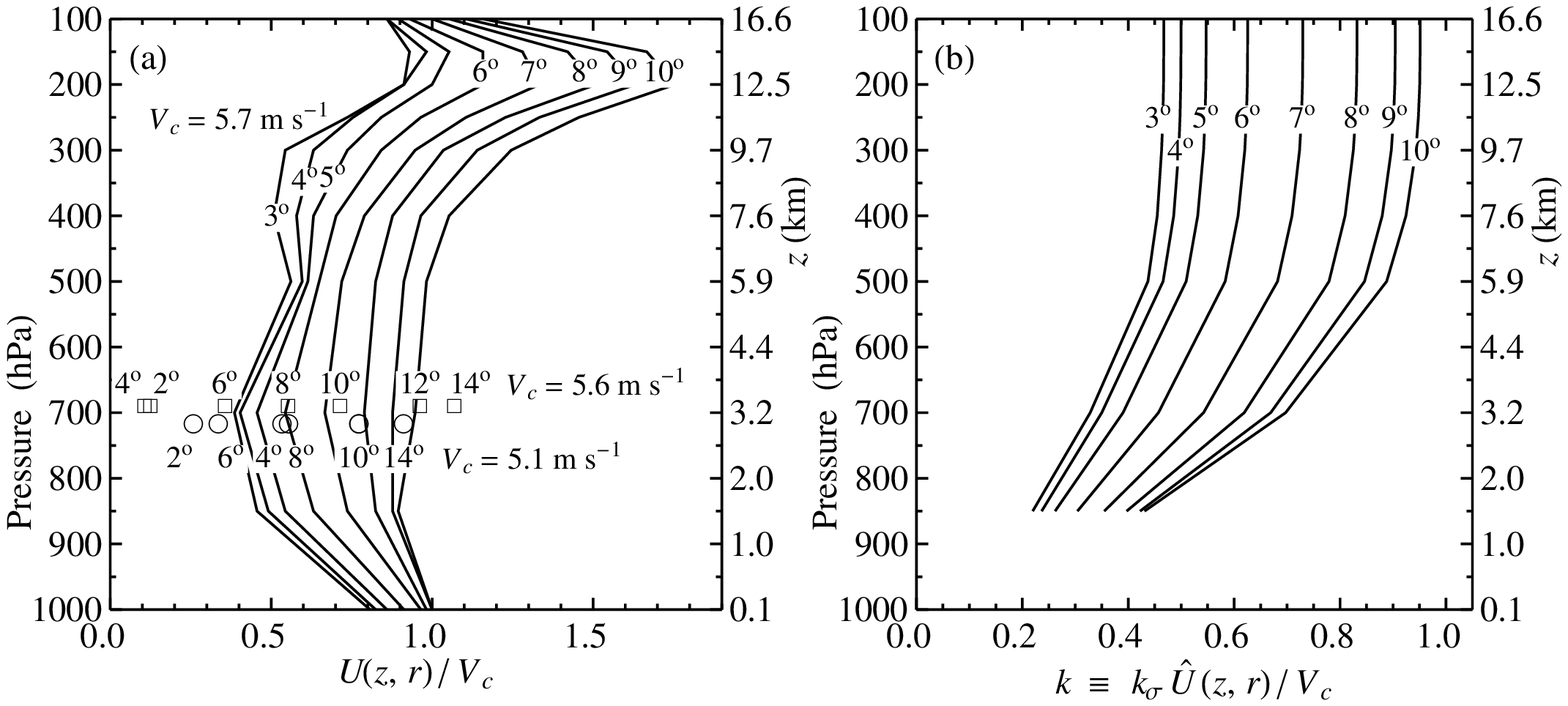}}
\caption{
Propagation velocity versus height $z$
at selected radial distances $r$ (shown in degrees latitude).
(a) Ratio of propagation velocity to translation velocity $U(z,r)/V_c$, solid curves --
data from Table~3 of \citet{franklin96} for Atlantic tropical cyclones;
symbols at 700~hPa -- data from Figs.~12 and 13 of \citet{george76} for
North Pacific tropical cyclones to the north (squares) and to the south (circles) of 20$^{\rm o}$~N; (b)
$k \equiv k_\sigma k_U $ (see Eq.~\ref{k}) for Atlantic cyclones from (a).
}
\label{figU}
\end{figure*}

There are few studies where
variation in local values of $U$ with height and distance from the cyclone center would be systematically assessed.
The limited available evidence indicates that propagation velocity $U = k_U V_c$ increases
with distance from the hurricane center. At $200$-$300$~km from the center $U$ is about 1.5-2~m~s$^{-1}$, or about one third
of a typical translation velocity $V_c$. It approaches or even exceeds $V_c$ at $r\sim 1000$~km (Fig.~\ref{figU}a).
This indicates that near its center the hurricane moves as a relatively rigid system:
for small $r$ the ring $r,r+dr$ moves at approximately the same mean velocity as the storm's inner circle.
With growing radial distance the storm's becomes less "rigid": the inner hurricane parts begin to move relative
to the surroundings with a considerable velocity. Another feature is a maximum of $U$ at the surface as compared to
the mid troposphere (Fig.~\ref{figU}a).

Mean propagation velocity $\hat{U}$ and radial velocity $\hat{V}_r$ in the inflow layer of height $z$ (weighted by water vapor content $q\rho$,
see Eq.~\ref{ks})
are
\beq\label{U}
\hat{U}(z,r) = \frac{\int_0^z U(z',r) q(z',r)\rho(z',r) dz'}{ \sigma(z,r)},
\eeq
\beq\label{Vr}
\hat{V}_r(z,r) = \frac{\int_0^z V(z',r) q(z',r)\rho(z',r) dz'}{ \sigma(z,r)},
\eeq
where $\sigma(z,r)$ is given by Eq.~(\ref{ks}).

For coefficient $k$ in Eq.~(\ref{Pa}) using Eqs.~(\ref{ks}) and (\ref{U}) we have
\beq\label{k}
k(z,r) \equiv k_U(z,r) k_\sigma(z) = \frac{\hat{U}(z,r)}{V_c} \frac{\sigma(z,r)}{\sigma_0(r)},\,\,\,k_U(z,r) \equiv \frac{\hat{U}(z,r)}{V_c}.
\eeq
Using the available data on $U$ (Fig.~\ref{figU}a) and $k_\sigma$ (Fig.~\ref{WVH}c)
we find that $k(z,r)$ varies from $0.2$ to unity for $300~{\rm km} \lesssim r \lesssim 1100$~km
and $1.5~{\rm km} \lesssim z \lesssim 17$~km (Fig.~\ref{figU}b). For a given radial distance there is little change of $k$ above 400~hPa, where
$k$ reaches its maximum. This reflects the absence of water vapor in the upper atmosphere -- whatever flows
in or out of the hurricane above $400$~hPa does not affect its moisture budget.

As we discussed in Section~\ref{sPm}, at the radius where the hurricane actually takes in the ambient water vapor,
mean radial velocity in the inflow layer should be related to propagation velocity as $\hat{V}_r$ = $\hat{U}/\pi$.
At this radius $k(z,r)$ (\ref{k}) should coincide with $k_r(z,r)$ defined as
\beq\label{kr}
k_r(z,r) \equiv  \pi \frac{\hat{V}_r(z,r)}{V_c}k_\sigma= \pi \frac{\hat{V}_r(z,r)}{V_c}\frac{\sigma(z,r)}{\sigma_0(r)}.
\eeq
Figure~\ref{figVr}b shows that for $200~{\rm km} \lesssim r \lesssim 900$~km
and $1.5~{\rm km} \lesssim z \lesssim 17$~km we have $0.1 \leqslant k_r(r,z) \leqslant 1$.
Where $k_r(z,r)$ and $k(z,r)$ actually coincide is discussed below (see Section~\ref{obs}).
We note that $q(z)$ and $\sigma(z)$ used to calculate $k_r$ and $k$ in Figs.~\ref{figVr}b and \ref{figVr}c correspond
to 20$^{\rm o}$N in September (see Fig.~\ref{WVH}b). Using other months and/or latitudes does not affect the resulting
values of $k_r$ and $k$ because of the approximate independence of $q(z)/q_s$ and $k_\sigma = \sigma(z)/\sigma_0$ of,
respectively, $q_s$ and $\sigma_0$ (Fig.~\ref{WVH}c).

\subsubsection{Evaporation}

\begin{figure*}[h!]
\centerline{\includegraphics[width=7cm]{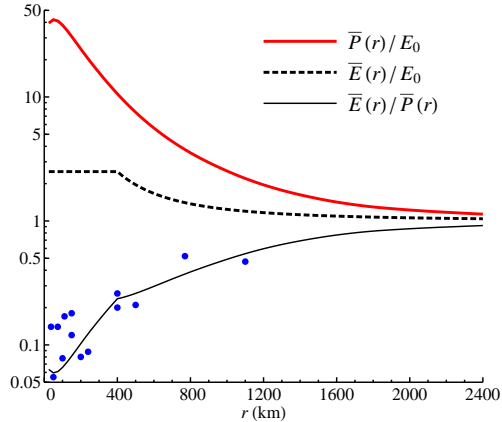}}
\caption{
Observed rainfall $\overline{P}(r)$ (Fig.~\ref{RR}d),
evaporation $\overline{E}(r) = k_E(r) E_0$ given by Eqs.~(\ref{kE1})-(\ref{kE2}) with $k_{E1}=2.5$ and $k_{E2} = 1$), both
in units of long-term evaporation $E_0 = 0.18$~mm~h$^{-1}$, and their ratio versus radial distance
$r$ from hurricane center. Circles are $\overline{E}/\overline{P}$ ratios for North Atlantic hurricanes from Table~1.
}
\label{figkE}
\end{figure*}

To specify coefficient $k_E$ that describes the input of evaporation
into hurricane rainfall in Eq.~(\ref{Pa}), we take into account the observed rainfall distribution $\overline{P}(r)$ (Fig.~\ref{RR}d).
Mean precipitation within a circle of $r_o = 400$~km is 1.9~mm~h$^{-1}$, or about ten times the long-term
evaporation mean, $\overline{P}(r_o) = 10E_0$ (Fig.~\ref{figkE}). On the other hand,
according to Table~1 and in agreement with Eq.~(\ref{rat}), we have $\overline{E}(r_o)/\overline{P}(r_o) = 0.25$.
Thus, we have $\overline{E}(r_o) = 2.5 E_0$. This means that evaporation within the outermost closed isobar of the hurricane
on average exceeds the long-term mean by two and a half times:
\begin{numcases}{k_E=}
k_{E1}, & if $r \leqslant r_o$    \label{kE1} \\
\displaystyle k_{E2} + (k_{E1}-k_{E2})\left(\frac{r_o}{r}\right)^2,    & if $r > r_o$   \label{kE2}
\end{numcases}
where $r_o = 400$~km, $k_{E1} = 2.5$ and $k_{E2} = 1$.

Outside the outermost closed isobar, $r > r_o$, local evaporation is assumed to be equal
to its long-term mean, $E(r) = k_{E2} E_0$, $k_{E2} = 1$. Then mean evaporation for $r > r_o$ is given
by $\overline{E}(r) \equiv k_E(r) E_0 = [k_{E1} E_0\pi r_o^2 + k_{E2} E_0 \pi (r^2 - r_o^2)]/\pi r^2$,
which is Eq.~(\ref{kE2}). Relationship (\ref{kE2}) captures the growth of evaporation-to-rainfall ratio with
increasing $r$ (Fig.~\ref{figkE}). For $r = 1000$~km we have $\overline{E}(r)/\overline{P}(r) \sim 0.5$ in agreement with
the results of \citet{jiang08} for Isidore (Table~1, Fig.~\ref{evap}).
Equation~(\ref{kE1}) may not be valid for $r \ll r_o$, but our results
to be obtained will pertain to larger $r > r_o$.

\subsection{Observational evidence}
\label{obs}

\begin{figure*}[h!]
\centerline{\includegraphics[width=12.9cm]{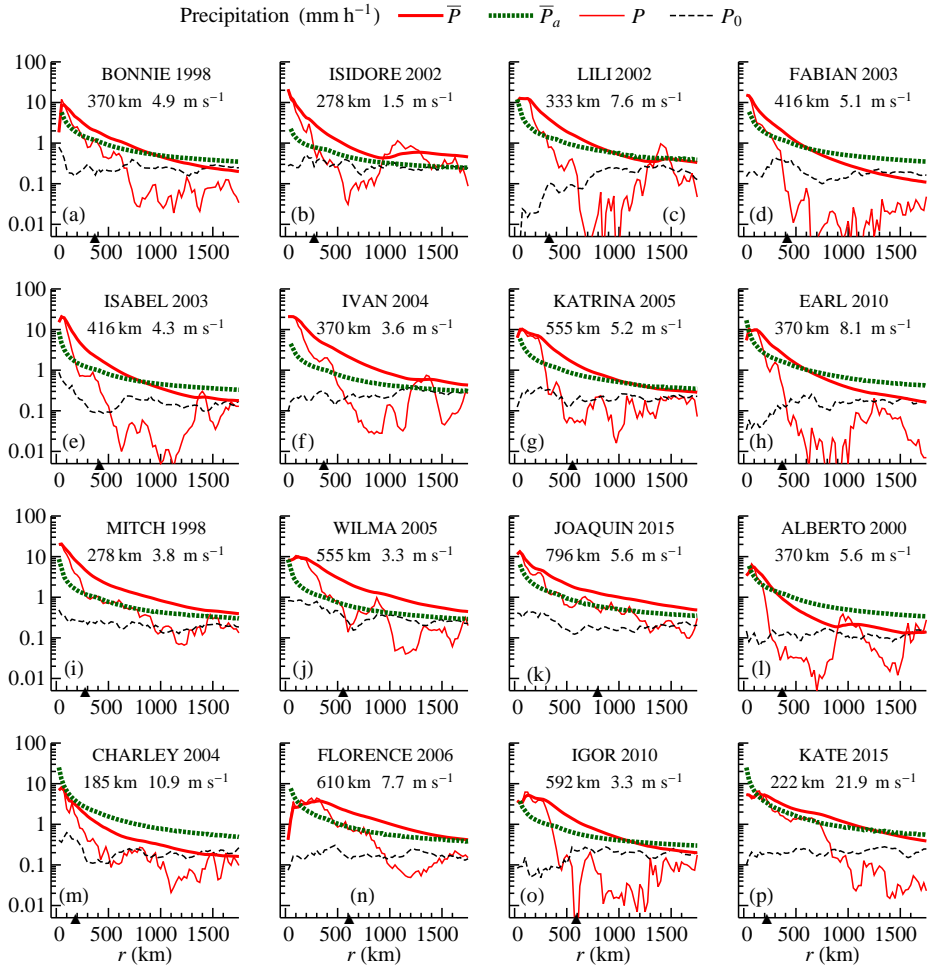}}
\caption{\label{Hurri2}
Observed $\overline{P}(r)$ versus available $\overline{P}_a(r)$ precipitation for $k = 0.5$, $k_{E1} = 2.5$ and $k_{E2}=1$ in Eqs.~(\ref{Pa})-(\ref{kE2})
for the same hurricanes as in Fig.~\ref{Hurri}, cf. Fig.~\ref{figPa}a.
Radius $r_o$ of the outermost closed isobar (marked with triangle on the horizontal axis) and translation velocity $V_c$
are from the EBTRK dataset.
}
\end{figure*}

We can now compare available and observed rainfall, $\overline{P}_a(r)$ (Eq.~\ref{Pa}) and $\overline{P}(r)$ (Fig.~\ref{RR}d).
We varied $k$ in Eq.~(\ref{Pa}) from $0.2$ to $1$ (as per Fig.~\ref{figU}b). For each $k$, for each $r$,
for each out of our 1551 hurricane observations we used its translation velocity $V_c$ calculated from the EBTRK dataset and its
water vapor distribution $\sigma_0(r)$ calculated from the MERRA dataset (see Methods and Fig.~\ref{WV}a). Putting
these values into Eq.~(\ref{Pa}) with $k_E$ given by Eqs.~(\ref{kE1})-(\ref{kE2}) we
obtained available rainfall $\overline{P}_a(r)$. Available versus observed rainfall in selected
hurricanes is shown in Fig.~\ref{Hurri2}.

\begin{figure*}[h!]
\centerline{\includegraphics[width=12.9cm]{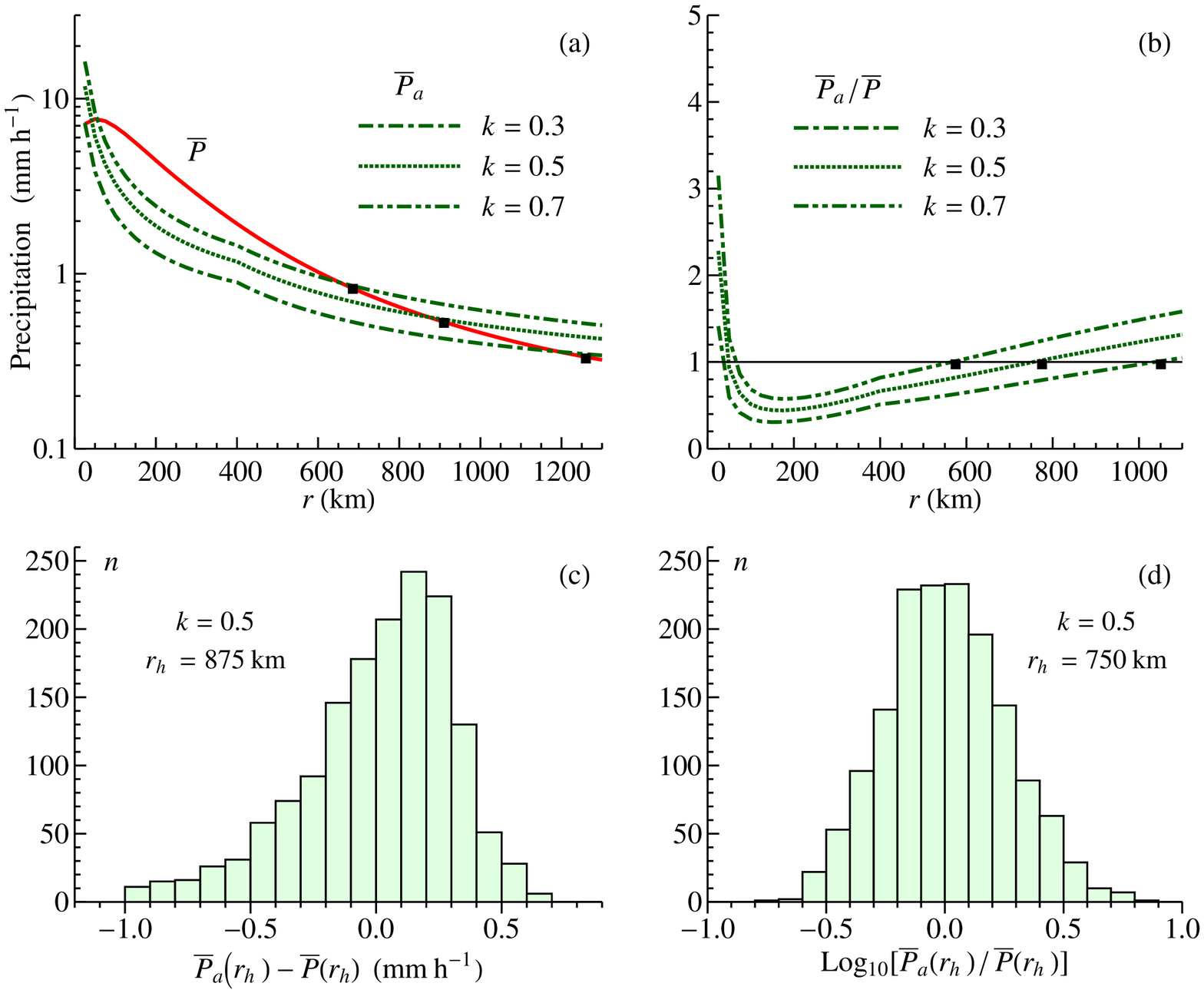}}
\caption{
Available versus observed precipitation for selected $k$ in Eq.~(\ref{Pa}). (a) Mean $\overline{P}_a(r)$ (mm~h$^{-1}$) averaged for each $r$
over all hurricane observations as compared to mean observed precipitation $\overline{P}(r)$ (mm~h$^{-1}$) (see Fig.~\ref{RR}d).
(b) Mean ratio $\overline{P}_a(r)/\overline{P}(r)$ averaged for each $r$ over all hurricane observations (log-normal distribution assumed).
(c,d) Frequency distribution of $\overline{P}_a(r_h)-\overline{P}(r_h)$ (mean $\pm$ one standard deviation $0.0016 \pm 0.87$~mm~h$^{-1}$) in (c)
and $\overline{P}_a(r_h)/\overline{P}(r_h)$ ($0.0063 \pm 0.25$) in (d) for $r_h$ corresponding to $k = 0.5$.
Filled squares in (a) and (b) indicate radius of self-sufficiency $r_h$ for each $k$.
}
\label{figPa}
\end{figure*}

We compared available and observed rainfall in our set
of $1551$ hurricane observations in two ways.
First, we averaged $\overline{P}_a(r)$ and $\overline{P}(r)$ separately
and then found $r = r_h$ where the two mean distributions coincide (Fig.~\ref{figPa}a).
Second, we averaged the ratio
$\overline{P}_a(r)/\overline{P}(r)$ for a given $r$
and found the maximum $r = r_h$ for which the mean ratio equals
unity (Fig.~\ref{figPa}b).
The first way yields a higher $r_h$ value; both values are reported below
as the range of $r_h$ estimates for each $k$.

\begin{figure*}[h!]
\centerline{\includegraphics[width=12cm]{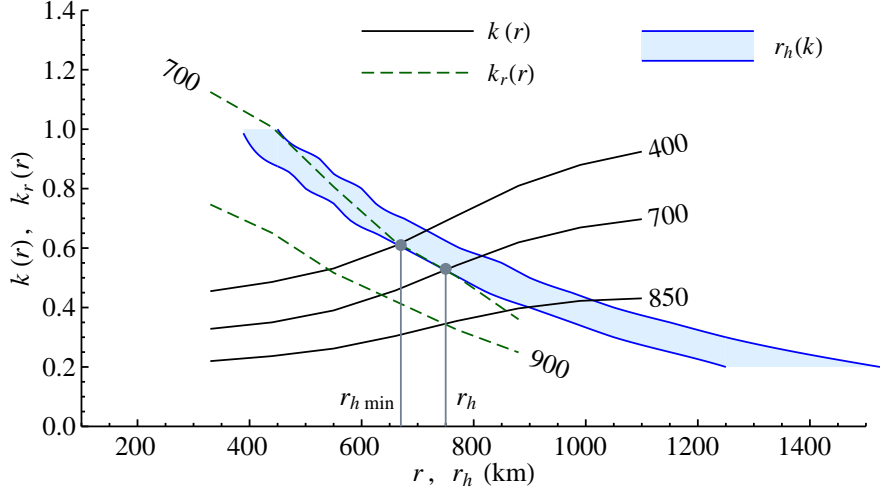}}
\caption{\label{res}
Self-sufficiency radius $r_h$ versus parameter $k$ in Eq.~(\ref{Pa}) (see text for details)
and parameters $k$ (Eq.~(\ref{k}), data from Fig.~\ref{figU}b)
and $k_r$ (Eq.~(\ref{kr}), data from Fig.~\ref{figVr}b of
\citet{miller58} for Atlantic hurricanes) versus radial distance $r$. Figures at curves denote height of the inflow layer
in hPa.
}
\end{figure*}

We find that the self-sufficiency radius $r_h$ grows with declining $k$ (Fig.~\ref{res}).
Inded, the value of $k$ reflects how fast the hurricane moves relative to the ambient air and how much of the available water vapor
it uses up (Eq.~\ref{k}). Thus, with decreasing $k$ available rainfall $\overline{P}_a(r)$ becomes smaller at all $r$, including $r = r_h$.
Since observed rainfall $\overline{P}(r)$ declines with increasing $r$, the now smaller $\overline{P}_a$
can equal $\overline{P}$ at a larger $r$ only. So $r_h$ is larger for a smaller $k$.

The minimum value of $r_h(k)=r_{h~\rm min}$ is constrained by the dependence of $k$ on $r$ (Fig.~\ref{res}).
Since propagation velocity grows with increasing $r$ (Fig.~\ref{figU}a), so does $k(r)$.
Thus, while for $k \to 1$ we find small values of $r_h$, these solutions are unrealistic: they would
require too large propagation velocities at small radii. Realistic solutions are defined by the
intersection of the curves $k(r)$ and $r_h(k)$ in Fig.~\ref{res}.
Since for any given $r$ the maximum value of $k(r)$ corresponds to the inflow layer of $400$~hPa (Fig.~\ref{figU}b),
we conclude that the minimum value of $r_h$ is $r_{h~\rm min} = 670$~km corresponding to the intersection
between $k(r)$ for $400$~hPa and the range of solutions $r_h(k)$ (Fig.~\ref{res}).
If the inflow layer is thinner and does not reach up to $400$~hPa, then minimal $r_h$ increases; for an inflow layer below $850$~hPa $r_h$
must exceed $900$~km (Fig.~\ref{res}).

Since, as we discussed in Section~\ref{sPm}, $\overline{P}_a(r)$ only gives the upper limit of rainfall available through motion,
it is essential to find out whether the hurricanes actually take in the moisture they sweep by their diameter $2r_h$.
If they do, mean radial velocity $\hat{V}_r(r)$ at $r = r_h$ must equal $\hat{U}(r)/\pi$. This requires that
three curves, $k_r(z,r)$ (Eq.~\ref{kr}), $k(z,r)$ (Eq.~\ref{k}) and $r_h(k)$ must intersect at one point in Fig.~\ref{res}.
They do indeed: for the inflow layer below $700$~hPa ($z\approx 3$~km)
we find that $k(r)$ and $k_r(r)$ intersect at $r = 750$~km. Here $k(r) = k_r(r) = 0.5$, which falls within the interval of estimated
$r_h(k)$ for $k=0.5$. Since above $700$~hPa the value of $k_r(r)$ grows little (see Fig.~\ref{figU}b), it follows from Fig.~\ref{res}
that there can be additional solutions for deeper inflow layers ($z \geqslant 3$~km) that correspond to $r_{h~\rm min} \leqslant r_h < 750$~km.

A hurricane's water budget can viewed as follows.
At $r = r_h \approx 700$~km the hurricane's inflow layer is at least $z_i = 3$~km thick (up to $700$~hPa).
This inflow contains about $70\%$ of the total moisture content in the atmospheric column: we have $k_\sigma = 0.7$
in Eq.~(\ref{Pa}) (see Eq.~(\ref{ks}) and Fig.~\ref{WVH}c).
The hurricane's mean propagation velocity in the inflow layer
is about $70\%$ of its translation velocity: $k_U = \hat{U}(z_i,r_h)/V_c = 0.7$ (Eq.~\ref{U}).
(Note that $k_U$ and $k_\sigma$ have coincided by chance as $k=k_U k_\sigma = 0.5$ in Fig.~\ref{res};
they are generally independent parameters.)
Mean radial velocity $\hat{V}_r$ at $r = r_h$ is related to propagation velocity
as $\hat{V}_r = \hat{U}/\pi$, which indicates that all water vapor the hurricane acquires by motion in the inflow layer
actually condenses and precipitates within it.
Thus, moving through the atmosphere with propagation velocity $\hat{U} = 0.7 V_c$ and consuming
all water vapor it encounters in the lower $3$~km, such a hurricane can sustain its rain observed
within the circle $r \leqslant r_h = 700$~km. Mean rainfall within area $r \leqslant r_h$ is about four
times the long-term mean (Fig.~\ref{figPm}).

\section{Hurricane water vapor budget and efficiency}
\label{eff}

The question of how the hurricane gathers its fuel -- water vapor -- is important
for finding constraints on hurricane power and for predicting hurricane frequency.
We have established that hurricanes can sustain rainfall by consuming {\it pre-existing}
atmospheric moisture and leaving the atmosphere drier than it was before the hurricane (Fig.~\ref{WV}c,f).
We now consider how this view relates to the commonly held alternative view of hurricanes
as steady-state thermodynamic systems that consume heat and moisture from the ocean \citep{emanuel86,emanuel91,ozawa15,kieu15}.

Kinetic power $W_K$  (W~m$^{-2}$) per unit hurricane area is commonly estimated as
$W_K = \rho C_D V^3$, where $V$ is air velocity and $C_D \sim 2\times 10^{-3}$
is surface exchange coefficient \citep{emanuel99}.
Viewing hurricane as a heat engine
consuming heat flux $Q_O$ from the ocean, its maximum kinetic power is estimated
as
\begin{equation}\label{car}
W_K = \rho C_D V^3 = \varepsilon_C Q_O,\,\,\,\varepsilon_C \equiv \frac{\Delta T}{T_s},
\end{equation}
where $\varepsilon_C$ is Carnot efficiency, $T_s$ is surface temperature, $\Delta T \equiv T_s - T_{out}$, $T_{out}$ is the
minimum temperature of hurricane air in the upper atmosphere \citep{ozawa15}. For
$T_s \approx 300$~K and $\Delta T$ between $75$~K and $105$~K \citep{demaria94}
we have on average $\varepsilon_C = 0.3$.

In the above analysis the assumption that heat influx $Q_O$ occurs concurrently with the hurricane is crucial.
This is because in theory the Carnot cycle efficiency relates heat input
to work (J); it does not relate heat influx to power output (J~s$^{-1}$). To obtain power from work
a characteristic time $\tau$ of the considered thermodynamic cycle must be known. If heat input and kinetic energy production
in the hurricane occur at the same time scale $\tau_Q = \tau_W$, i.e. concurrently, then hurricane power $W_K$ can be constrained
by Carnot efficiency (Eq.~\ref{car}). If, on the other hand, heat input occurs on a much longer time scale than
kinetic energy production in the hurricane, $\tau_Q \gg \tau_W$, then Eq.~(\ref{car}) will constrain the {\it mean kinetic
energy production rate} observed during time $\tau_Q$. This mean rate will be by a factor of $\tau_W/\tau_Q \ll 1$ smaller
than the actual kinetic power output in a hurricane rendering Eq.~(\ref{car}) irrelevant.

An alternative approach is to consider the partial pressure of water vapor $p_v$
(J~m$^{-3}$) as a store of potential energy equal to $p_v/N_v = RT$ (J~mol$^{-1}$)
\citep{mgpla09,pla11,dhe10,m13}. This energy is released when water vapor condenses, which
can occur at an arbitrarily high rate proportional to the vertical velocity of the ascending air.
Since potential energy can be converted to kinetic energy with
efficiency equal to unity, this approach predicts that kinetic power output
is proportional to rainfall:
\begin{equation}\label{WK}
W_K = \overline{P}RT,
\end{equation}
where $\overline{P}$ (mol H$_2$O m$^{-2}$ s$^{-1}$) is mean rainfall within the hurricane and $T$ is the mean temperature of
condensation \citep{pla11}.

And, indeed,
Eq.~(\ref{WK}) explains the observed average dependence of maximum velocity $V_{\rm max}$ on rainfall in
cyclones with $V_{\rm max} \geqslant  30$~m~s$^{-1}$ \citep{pla15}.
On the other hand, the Carnot cycle view on hurricanes would be able to persist -- as it does -- only
if Eq.~(\ref{car}) agreed with observations at least for some of the most intense hurricanes.
How can the two alternative approaches give similar results?

Assuming that evaporation accounts for three quarters of total heat influx from the ocean \citep{jaimes15}
we have for $Q_O$ in Eq.~(\ref{car})
\begin{equation}\label{QO}
Q_O = \frac{4}{3} \overline{E} L,
\end{equation}
where $L = 45$~kJ~mol$^{-1}$ is latent heat of evaporation.

Using Eq.~(\ref{QO}) in Eq.~(\ref{car}) we observe that the two governing equations
for the two alternative approaches, Eq.~(\ref{car}) and Eq.~(\ref{WK}),
would yield the same estimate of hurricane power $W_K$ if
\beq\label{coin}
\frac{RT}{L} = \frac{4}{3} \frac{\overline{E}}{\overline{P}}\varepsilon_C .
\eeq

It so happens that for intense hurricanes where $\overline{E}/\overline{P}$ ratio is low, Eq.~(\ref{coin})
is approximately fulfilled. For $T \approx T_s - 15~K$, $T_s = 300$~K, $\varepsilon_C = 0.3$ and
$\overline{E}/\overline{P}= 0.13$ (as we estimated for hurricane Isabel in Section~2) we have
\beq\label{coinnum}
\frac{RT}{L} = 0.053 \approx \frac{4}{3} \frac{\overline{E}}{\overline{P}} \varepsilon_C = 0.052.
\eeq

This analysis explains why, unlike Eq.~(\ref{WK}), which is universally valid \citep{pla15}, Eq.~(\ref{car}) can match observations only in
intense hurricanes. In less intense storms, as we discussed in Section~2, condensation in the rising air
is incomplete, and the evaporation-to-rainfall ratio in Eq.~(\ref{coinnum}) increases.
In the result, the right-hand part of the approximate equality in Eq.~(\ref{coinnum}) exceeds $RT/L$.
Thus, in less intense hurricanes with a larger evaporation-to-rainfall ratio, the Carnot constraint (\ref{car}) does not match
observations but should exceed the observed power (\ref{WK}).
Conversely, one can predict that the actual hurricane power described by Eq.~(\ref{WK}) can exceed the Carnot constraint
(\ref{car}) in those hurricanes where the evaporation-to-precipitation ratio is small. As we discussed
in Section 2, this ratio should be small in those hurricanes where surface air temperature decreases towards the hurricane
center -- thus contributing a negative third term to the right-hand side of Eq.~(\ref{dg}). Accordingly, it was hurricane Isabel
where air temperature was dropping by 2~K as the air was approaching the windwall, that was reported to have exceeded
the maximum potential intensity estimated from Eq.~(\ref{car}) \citep{montgomery06}.

We conclude that the approximate agreement between Eqs.~(\ref{WK}) and (\ref{car}) for intense hurricanes can be traced to an accidental numerical
agreement between unrelated variables (Eq.~\ref{coin}). This coincidence
explains how Eq.~(\ref{car}) can match selected observations despite hurricanes not in fact behaving as Carnot heat engines.
Indeed, accounting for the atmospheric moisture source, the ratio between kinetic power and total latent heat flux $Q_{in} = \overline{P}L$ is
$W_K/Q_{in} = R T/ L \approx (1/5) \varepsilon_C \ll \varepsilon_C$.
It follows that the maximum observed hurricane power cannot be explained by viewing the hurricane
as a thermodynamic cycle converting external heat to work at Carnot efficiency.

\section{Conclusions}

Most rainfall within the outermost closed isobar of a hurricane is due to imported moisture rather than local evaporation (Table~1 and
Figs.~\ref{hur} and \ref{evap}). Here we have shown that while evaporation can be estimated
from the observed radial distribution of pressure, relative humidity and temperature in the boundary layer,
such estimates are insufficient to clarify the origin of imported moisture (Section~\ref{sec2}).

We then considered rainfall and moisture distribution
up to $3000$ km from the hurricane center using TRMM and MERRA data. For each out of $1551$ analyzed observations
of North Atlantic hurricanes in $1998$-$2015$ we defined reference distributions of rainfall and atmospheric moisture
characterizing the hurricane environment in those years when the hurricane was absent.
By comparing rainfall and moisture distributions in hurricanes to their seventeen years' hurricane-free means
we established that hurricanes leave a dry footprint of suppressed rainfall in their wake. At distances $r > r_d \sim 500$-$600$~km from the center
rainfall is reduced by up to 40\% compared to what it is on average in hurricane's absence
(Figs.~\ref{rmin} and \ref{Pm}). This pattern was not previously established in quantitative terms since
most studies of hurricane-related rainfall focused on a smaller area of up to $300$-$400$ km from the center
\citep[e.g.,][]{lonfat04}. The inner radius $r_d$ of hurricanes' dry footprint
corresponds to the so-called clear sky moat that often surrounds tropical cyclones at $440$-$660$~km
from the storm's center \citep{frank77}.

We also showed that rainfall in hurricanes typically remains considerably larger than the sum mean evaporation
up to $3000$~km from the hurricane center. This indicates that a hurricane cannot be viewed as
a steady-state system, whereby water vapor provided by {\it concurrent} evaporation
across a few thousand kilometers could be concentrated to the smaller rainfall area of a hurricane (Section~\ref{r&e}).

We offered a novel proposal for how hurricanes capture moisture: {\it pre-existing}
atmospheric water vapor stores are consumed as the storm moves through the atmosphere.
Rainfall available to the hurricane due to its movement is directly proportional to movement velocity $U$ (Eq.~\ref{Pm}).
We found that radius $r_h$ at which a moving hurricane becomes self-sufficient with respect to moisture
is about $r_h\sim 700$~km. This radius of self-sufficiency exceeds both the radius $r_o \sim 400$~km of
the uttermost closed isobar and the radius $r_d \sim 500$-$600$~km of hurricane's dry footprint: $r_h > r_d >r_o$ (Fig.~\ref{figPm}).
This means that huricane's movement provides for observed rainfall in the entire area $r \leqslant r_d$ where the hurricane-related rainfall
is elevated above the long-term mean characterizing a hurricane-free environment. Capturing its nearby water vapor by motion,
such a hurricane does not need to concentrate moisture from far away.

In the present study we introduced general relationships and key parameters relating rainfall to motion.
In particular, we proposed that propagation velocity $U$, i.e. storm's velocity relative to that of the surrounding air flow,
is key for the cyclone's water vapor budget.
So far propagation velocity has not been studied systematically as dependent
on radial distance and altitude. Many studies focused on mean values of $U$ for the entire troposphere averaged
across large radial distances \citep{chan05}.  We noted that
the cyclone's water budget is determined by the mean value of $U$ in the inflow layer at around $r = r_h \sim 700$~km (Fig.~\ref{figU}b).
This parameter, and especially its relationship to radial velocity (see Fig.~\ref{res} and Section~\ref{obs}),
apparently merits more extensive investigations. For example,
according to \citet[][see their Table~4]{george76}, more intense cyclones appear to have a larger $U$ than less intense ones.
In our analysis the three most intense hurricanes have their available rainfall below
observed rainfall at all $r$ (Fig.~\ref{Hurri2}i-k). These hurricanes also  having a higher than
average $U$ could explain this pattern.

\begin{figure*}[h!]
\centerline{\includegraphics[width=12.9cm]{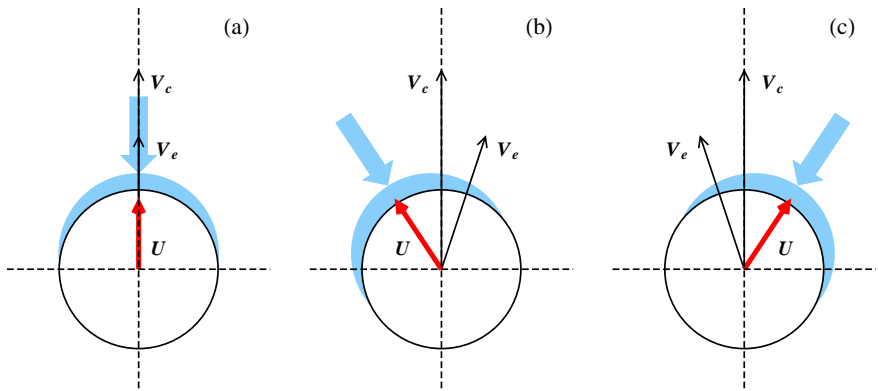}}
\caption{\label{figdir}
Predicted rainfall asymmetry (more rain is indicated by the blue area, cf. Fig.~\ref{hur2}b) as dependent on propagation velocity $\mathbf{U} = \mathbf{V}_c - \mathbf{V}_e$ in the inflow layer
for different orientations of translation velocity $\mathbf{V}_c$ of the cyclone relative to velocity $\mathbf{V}_e$
of the environmental flow. Blue arrow indicates the dominant direction of moisture import into the cyclone
associated with its movement.
}
\end{figure*}

In our analysis we used the approximation of radial symmetry. However,
cyclone motion relative to the ambient air must introduce asymmetries in the rainfall
distribution. These are to be further investigated.
According to Fig.~\ref{hur2}, the rainfall asymmetry should be related
to the direction of propagation velocity $\mathbf{U}$ relative to the storm motion.
The few available data indicate that in the lower layer of up to $700$~hPa tropical cyclones move approximately colinearly
with the environmental flow albeit at a faster speed \citep[see, e.g., Fig.~7 of][]{chan82}.
This means that in the frame of reference co-moving with the storm the environmental winds blow against
the moving storm: area $S$ in Fig.~\ref{hur2}b is located in the two
front quadrants of the moving storm (Fig.~\ref{figdir}a). This moisture influx should produce more rain ahead of the cyclone center.
This agrees with the findings of \citet{lonfat04} who established that
in tropical cyclones there is more rainfall in the front two quadrants and less rainfall in the rear two quadrants of the moving storm.

If the hurricane moves to the right (left) of the low-level environmental flow, then water vapor is predominantly
imported into the front left (right) quadrant, producing more rain there (Fig.~\ref{figdir}b,c).
For eastward moving Atlantic cyclones \citet{dong86} indicated that
while weakening storms tend to move to the left of the low-level environmental flow (Fig.~\ref{figdir}b),
intensifying storms as well as hurricanes tend to move to the right of it (Fig.~\ref{figdir}c).
This is consistent with the earlier findings of \citet[][Fig.~7]{miller58}
whereby in the lower 1-3~km radial velocity relative to the hurricane center is maximum in
the front right quadrant. These differences may explain the variation in rainfall
asymmetries between weak and intense storms \citep[see][]{lonfat04}.
The relationships between rainfall asymmetries and the vertical wind shear
between the $850$~hPa and $200$~hPa investigated by \citet{chen06} may also reflect the dependence
of rainfall on propagation velocity $U$, since the latter changes both its direction and magnitude
with altitude \citep[see, e.g.,][their Fig.~2]{dong86}.

Asymmetries in the spatial distribution of ambient water vapor should be equally important
to the water vapor budget. In our calculations we used the mean amount $\sigma_0(r)$ of water
vapor present in a hurricane-free environment. If, however,
hurricanes predominantly move in the direction of a higher water vapor content this would mean that $k$ in
Eq.~(\ref{Pa}) is higher and the radius of self-sufficiency $r_h$ is lower than we estimated. According to Fig.~\ref{hur2},
what matters most is the water vapor content available to the cyclone in the direction of its propagation velocity.

The formulated view of the hurricane as an open system which maintains itself by continuously consuming atmospheric moisture
as it moves explains the hurricane moisture budget and generally clarifies the nature of hurricanes
as steady-state systems. In particular, \citet{smith14} indicated that the source of angular momentum for
a steady-state hurricane remains unknown. That is, if we consider the air moving along path $\rm AB$ in
Fig.~\ref{hur}, the angular momentum of this air is known to diminish with decreasing $r$. Since
in the upper atmosphere angular momentum is approximately conserved as the air moves along path $\rm BCD$,
the question posed by \citet{smith14} was how, in a steady-state, the air gains angular momentum
as it descends from $\rm D$ to $\rm F$. Our view of the hurricane
as an open system identifies such a source: it comes with external air.
Having left the hurricane, the air does not return (Fig.~\ref{hur2}a).
In this framework, in the reference frame of the Earth the hurricane is not a steady-state circulation.
It would vanish if its motion relative to the air discontinues or if it encounters a dry area.

While the fact that rainfall greatly exceeds evaporation in intense cyclones has been long recognized, it has also been
maintained that, compared to evaporation, rainfall by itself does not play a significant role in cyclone formation
\citep{malkus60,emanuel91,gamache93,braun06,yang11,huang2014}. This
view stemmed from the observed relationship between the fall in surface pressure $\Delta p$ (key for hurricane development)
and water vapor increment in surface air $\Delta \gamma$. This relationship, usually expressed in terms of equivalent potential temperature,
is conventionally interpreted as a cause-effect relationship: no evaporation, no pressure fall, no hurricane \citep{malkus60,holland97}.
We note, however, that this relationship is similar to that between
surface pressure and temperature gradients which we have debated elsewhere \citep{jcli15}. As we previously showed,
such relationships are based on one key parameter -- the isobaric height.
Its magnitude is derived from observations and cannot prove cause-effect relationships
among the variables in question. In other words,
while it is commonly observed that the drop in surface pressure in intense hurricanes is accompanied by an increase in the moisture
content $\gamma$ of the surface air, this evaporation may as well be a result as well as a cause of the pressure change.
Despite its long history, we find no evidence to indicate that the
pressure gradients observed in intense cyclones could develop in the presence of intense evaporation without large-scale
atmospheric moisture supplies (and related rainfall).
But, in keeping with our interpretations, there is accumulating evidence to the contrary.

Intensity forecasts for tropical cyclones have been shown to be sensitive to the accurate representation
of rainfall \citep{krishnamurti93,krishnamurti98}. Recent observations and modelling studies indicate a correlation between cyclone
intensification and atmospheric moisture input \citep{fritz14,ermakov14,ermakov15}. Our theoretical framework indicates that
hurricane intensity is directly proportional to rainfall (Eq.~\ref{WK}). \citet{sabuwala15} established this proportionality empirically.
Here we have also shown that consideration of the atmospheric moisture supply and associated rainfall is key for
constraining hurricane intensity. We thus call for a broader re-assessment of the role
of the hurricane's moisture dynamics in storm development and maintenance.


{\bf Acknowledgments.}
We thank B. Jaimes, L.K. Shay and E.W. Uhlhorn for providing us with the data of their Fig.~11
and M. Bell, M. Montgomery and K. Emanuel for providing us with the data of their Fig.~15.
MERRA and TRMM data have been provided by the Global Modeling and Assimilation Office at NASA Goddard Space
Flight Center through the NASA GES DISC online archive
(http://mirador.gsfc.nasa.gov). This work is partially supported by Russian Scientific Foundation
Grant~14-22-00281, the University of California Agricultural Experiment Station, the Australian Research Council project DP160102107 and
 the CNPq/CT-Hidro - GeoClima project Grant~404158/2013-7.

\nobreak

\bibliography{met-refs}

\end{document}